\documentclass[english, twocolumn, 10pt, aps, superscriptaddress, floatfix, showpacs, prl citeautoscript,amsmath]{revtex4-1}

\usepackage{graphicx}
\usepackage{bm}
\usepackage{braket}
\usepackage{units}

\begin{document}

\title{Erasing odd-parity states in semiconductor quantum dots coupled to superconductors}

\author{Z. Su}
\affiliation{Department of Physics and Astronomy, University of Pittsburgh, Pittsburgh PA, 15260, USA} 
\author{R. \v{Z}itko}
\affiliation{Jo\v{z}ef Stefan Institute, Jamova 39, Ljubljana, Slovenia \\
Faculty of Mathematics and Physics, University of Ljubljana, Jadranska 19, Ljubljana, Slovenia}
\author{P. Zhang}
\affiliation{Department of Physics and Astronomy, University of Pittsburgh, Pittsburgh PA, 15260, USA} 
\author{H. Wu}
\affiliation{Department of Physics and Astronomy, University of Pittsburgh, Pittsburgh PA, 15260, USA} 
\author{D. Car}
\affiliation{Eindhoven University of Technology, 5600 MB, Eindhoven, The Netherlands}
\author{S.R. Plissard}
\affiliation{LAAS CNRS, Universit{\'e} de Toulouse, 31031 Toulouse, France}
\author{S. Gazibegovic}
\affiliation{Eindhoven University of Technology, 5600 MB, Eindhoven, The Netherlands}
\author{G. Badawy}
\affiliation{Eindhoven University of Technology, 5600 MB, Eindhoven, The Netherlands}
\author{M. Hocevar}
\affiliation{Univ. Grenoble Alpes, CNRS, Grenoble INP, Institut N\'eel, 38000 Grenoble, France}
\author{J. Chen}
\affiliation{Department of Physics and Astronomy, University of Pittsburgh, Pittsburgh PA, 15260, USA} 
\affiliation{Department of Electrical and Computer Engineering, University of Pittsburgh, Pittsburgh, PA 15261, USA} 
\author{E.P.A.M. Bakkers}
\affiliation{Eindhoven University of Technology, 5600 MB, Eindhoven, The Netherlands}
\author{S.M. Frolov}
\affiliation{Department of Physics and Astronomy, University of Pittsburgh, Pittsburgh PA, 15260, USA}

\date{\today}

\begin{abstract}

Quantum dots are gate-defined within InSb nanowires, in proximity to NbTiN superconducting contacts. As the coupling between the dot and the superconductor is increased, the odd-parity occupations become non-discernible (erased) both above and below the induced superconducting gap. Above the gap, conductance in the odd Coulomb valleys increases until the valleys are lifted. Below the gap, Andreev bound states undergo quantum phase transitions to singlet ground states at odd occupancy. We observe that the apparent erasure of odd-parity regimes coincides at low-bias and at high-bias. This observation is reproduced in numerical renormalization group simulations, and is explained qualitatively by a competition between Kondo temperature and the induced superconducting gap. In the erased odd-parity regime, the quantum dot exhibits transport features similar to a finite-size Majorana nanowire, drawing parallels between even-odd dot occupations and even-odd one-dimensional subband occupations.

\end{abstract}

\maketitle

The prospect of topological quantum computing has motivated recent studies of charge parity, which is an observable that can distinguish the states of a Majorana quantum bit \cite{sarma2015majorana}. Experiments have demonstrated that electrons can be added to small superconducting islands either in pairs or one at a time, depending on the interplay of energy scales of the system \cite{TuominenPRL92, eilesPRL93, LafargePRL93, AlbrechtNature2016}. Quantum jumps of parity have been observed in superconducting circuits \cite{janvierscience15, haysprl18}. In quantum dots coupled to superconductors, a quantum phase transition (QPT) between odd- and even-parity ground states has been investigated theoretically \cite{matsuura1977,satori1992,yoshioka2000,choi2004josephson, oguri2004josephson,bauer2007,karrasch2008} and experimentally \cite{EichlerPRL07, SandJespersenPRL07, BuizertPRL07, GroveRasmussenPRB09, DeaconPRL10, KanaiPRB10, first_ABS,  MaurandPRX12, ChangPRL13, KumarPRB14, LeeNatnano2014, JellinggaardPRB16, lee2017prb}.

In this paper, we study the transition from the closed to the open regime (i.e., weak hybridization to strong hybridization regime) in a quantum dot defined in a semiconductor nanowire with superconducting contacts. We demonstrate how the transport signatures of odd-parity quantum-dot states are erased in the open dot regime. At high source-drain bias, we observe a transition from well-defined Coulomb diamonds to a conductance modulation pattern in which odd Coulomb valleys rise in conductance faster than the even valleys. We explain this based on Kondo physics playing a role at stronger coupling. At low bias, we observe Andreev bound states (ABS) that cross zero energy at every change in ground state parity from even to odd, while in the open regime the ABS do not cross zero bias. This is the manifestation of a well-studied QPT upon which the system always stays in a many-body singlet ground state. 

Interestingly, we notice that the ABS QPT at low bias coincides with the lifting of the odd Coulomb valleys at high bias. We furthermore reproduce this observation qualitatively in the simulation of the Anderson impurity model using the numerical renormalization group (NRG). We argue that this coincidence is expected because both high-bias and low-bias features are determined by the competition between the Kondo temperature and the induced gap. 

In the same system we furthermore find similarities between the evolution of ABS and the emergence of Majorana bound states (MBS). Specifically, the phase diagram of the ABS QPT in gate voltage and magnetic field is reminiscent of the phase diagram predicted for the topological superconducting phase, if even-odd dot occupations are replaced with even-odd subband occupations. This highlights the persistent need to identify truly unique signatures of MBS and to deeper study the related regime of trivial ABS.

\begin{figure}[ht]
  \includegraphics[width=\columnwidth]{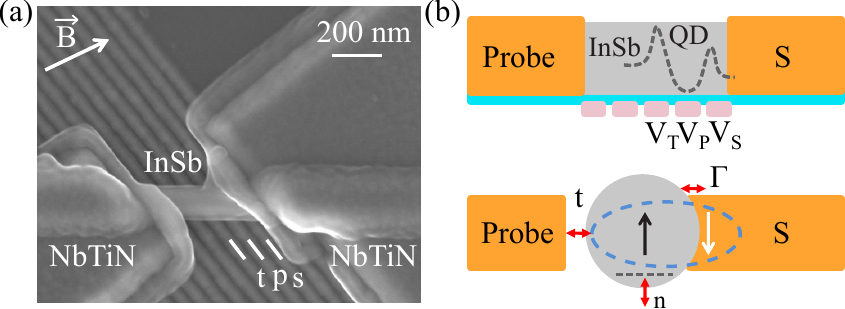}
  \caption{
(a) Scanning electron micrograph of the device studied in the main text: InSb nanowire in contact with superconducting NbTiN contacts. The arrow indicates the direction of the applied magnetic field, gates $t$, $p$, $s$ are labeled. 
(b) Side-view schematic of the device: superconducting contacts (orange), nanowire (grey), gate electrodes (pink), dielectric (blue), dashed line is an approximate potential created by gates to form a quantum dot (QD). (c) Superconductor on the right couples to the dot with hybridization $\Gamma$, while the superconductor on the left acts as a tunnel probe (t).
 }
 \label{fig1}
\end{figure}

InSb nanowires are grown using metalorganic vapor phase epitaxy (MOVPE) and transferred onto metallic gate patterns which are lines with \unit[60-80]{nm} center-to-center pitch. The gates are covered by 10 nm of HfO$_2$ dielectric. NbTiN superconducting contacts are fabricated on top of the nanowire. The device in Fig.\ref{fig1}(a) is highly tunable: previous reports on the same device demonstrated that it can be used to set up a quantum dot near the left or the right superconductor, as well as a double dot \cite{su_andreevmolecule, SuPRL18}. Here a single quantum dot is defined using gates $t$, $p$, $s$ near the right superconductor. The barrier controlled by gate $t$ is set near pinch off, defining the left contact as the tunnel probe. Even though the probe contact is a superconductor, the soft gap effect introduces finite density of states at the Fermi level. This makes the low bias spectra similar to those obtained with a normal probe \cite{SuPRL18}. The barrier controlled by the voltage $V_S$ on gate $s$ tunes $\Gamma$, the coupling to the right superconductor  (Fig.\ref{fig1}(b)). The voltage on gate $p$, $V_P$, primarily controls the dot chemical potential $\mu$ or the number of electrons $n$. The minimal 2-terminal resistance of the device is 4 k$\Omega$. Measurements are performed in a dilution refrigerator with a base temperature of \unit[40]{mK}. 

\begin{figure}[ht]
  \includegraphics[width=\columnwidth]{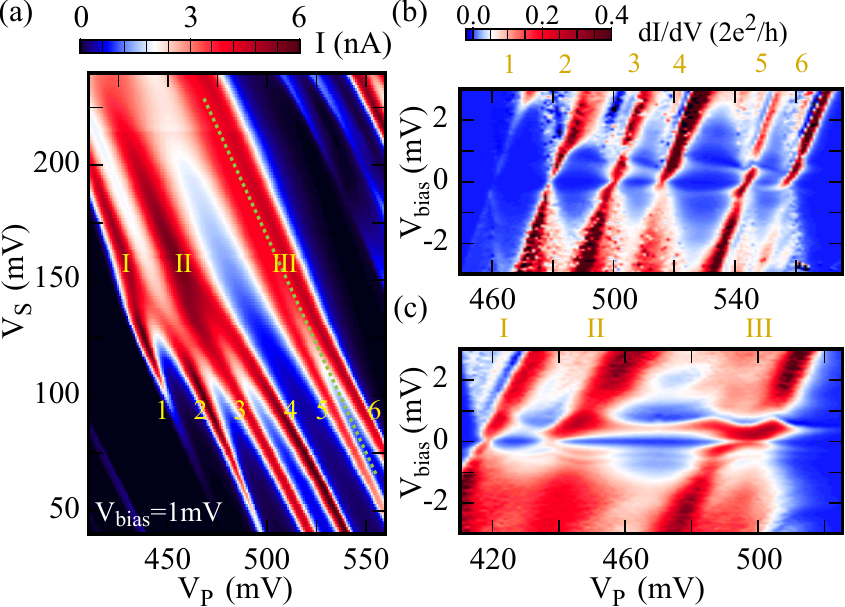}
  \caption{
(a) Current across the device at bias voltage $V_\mathrm{bias}=\unit[1]{mV}$ versus gate voltages $V_P$ and $V_S$. The arabic numerals $1-6$ denote the Coulomb peaks in the weak-coupling regime, the latin numerals I, II, III indicate the broad resonances in the strong-coupling regime. Current along the dashed green-line plotted in Fig.~3(g).
(b) Finite-bias spectroscopy at weak-coupling shows six Coulomb diamonds, and in-gap states forming loops at low bias, $V_{bias} \lesssim \unit[0.4]{mV}$. 
(c) Finite-bias spectroscopy at strong-coupling shows  broad (2e) resonances at high bias and anticrossing sub-gap features at low bias.
 }
 \label{fig2}
\end{figure}

\begin{figure}[ht]
  \includegraphics[width=\columnwidth]{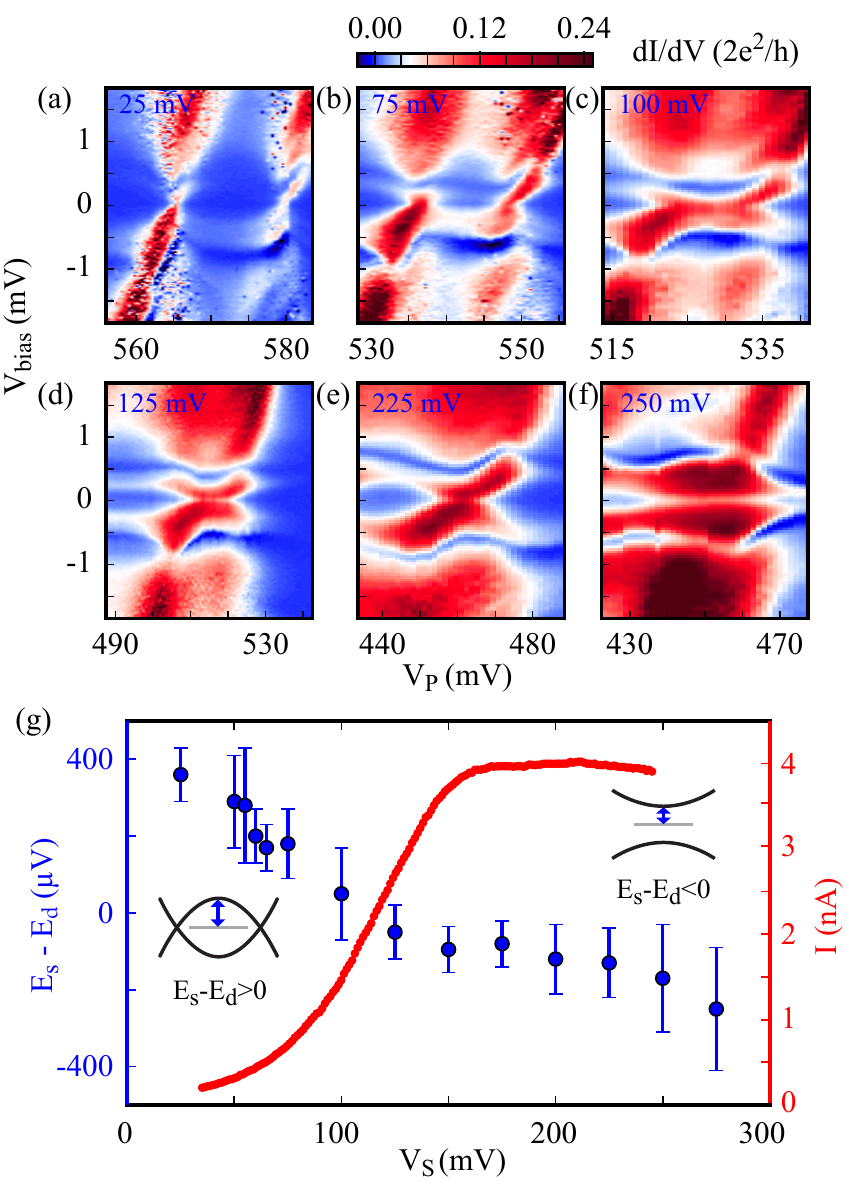}
  \caption{
(a-f) Evolution of the finite-bias spectra of the odd Coulomb valley 'III' from weak to strong coupling. The gate voltage $V_S$ is indicated in blue in the upper left of each panel.
(g) Comparison of the singlet-doublet energy difference (blue points), and the finite-bias current in valley 'III' (red line).
 }
 \label{fig3}
\end{figure}

The key experimental observations of the paper are presented in Figs.~\ref{fig2} and \ref{fig3}. For $V_S < \unit[100]{mV}$ we observe a sequence of Coulomb peaks by tuning $V_P$; we label these peaks 1 through 6 in Figs. \ref{fig2}(a,b). For $V_S > \unit[100]{mV}$ Coulomb peaks are seen continuously merging pairwise into broader resonances labeled 'I', 'II', 'III' in Figs.~\ref{fig2}(a, c). Fig. \ref{fig2}(a) is obtained under a source drain bias voltage $V_{bias}=1 mV$, above the apparent induced gap. It shows that current in every other Coulomb valley increases with higher $V_S$, i.e. the odd valleys are being lifted. The effect of transitioning from $1e$ to $2e$ conductance pattern in open quantum dots can be qualitatively understood as follows. By making the tunneling barrier more transparent, the effect of the Coulomb interaction on the dot is suppressed: at very high hybridization the Coulomb energy $U$ becomes irrelevant and the system behaves as if no additional energy were required to add a second electron to the same quantum dot orbital state \cite{steele_unpublished}. 

At lower biases, a set of horizontal resonances near $V_{bias} = \unit[0.4]{mV}$ is observed (Figs.~\ref{fig2}(b,c)).  In Fig.~\ref{fig2}(b), the horizontal resonances appear to coexist with Coulomb resonances 1-6. Such resonances have been reported in quantum dots with superconducting contacts and are either related to co-tunneling enhanced by high density of states at the gap edge \cite{GroveRasmussenPRB09}, or related to ABS in the dot.
In Fig.~\ref{fig2}(c),  at higher $V_S$, when the quantum dot is stronger coupled to the right superconductor, the horizontal resonances are still present, but the Coulomb peaks are absent. Instead, at zero bias conductance exhibits a local minimum throughout the range of $V_P$. At high bias, Coulomb diamonds are replaced with three broader resonances I,II,III.

The detailed evolution of low-bias superconductivity-related resonances as a function of $V_S$ is shown in Figs.~\ref{fig3}(a)-(f), which is focusing on the region labeled 'III' in Fig.~\ref{fig2}. For low 
$V_S$ (panels (a)-(b)) the resonances form a loop around zero bias. The loop shrinks with increasing $V_S$ until the two zero-bias crossings merge around $V_S = \unit[100]{mV}$ (panel (c)). For higher $V_S$, the horizontal resonances at positive and negative biases exhibit an anticrossing, its level repulsion growing with more positive $V_S$ (panels (d),(e),(f)). This transition from loop-like to anticrossing-like resonances has been previously studied as a manifestation of the singlet-doubled QPT at odd quantum dot occupancy \cite{martin2011josephson,Luitz:2012jz,lee2017prb,meden2018}. Supplemental materials contain comprehensive data on the evolution from the closed to open dot regime over wider ranges of $V_{bias}$ and $V_P$.

In Fig.~\ref{fig3}(g) we correlate the lifting of the odd Coulomb valley 'III' with the singlet-doublet ABS QPT. To identify the valley-lifting point, we plot in solid line a trace of current along a green dashed line in Fig.~\ref{fig2}(a) at $V_{bias} = \unit[1]{mV}$. When $V_S$ is low, the current is near zero because of Coulomb blockade. As $V_S$ is increased, the dot becomes open and the current starts to increase and finally reaches $\unit[4]{nA}$. 

To identify the singlet-doublet transition, we plot the bias voltage of the lowest positive resonance in the center of region 'III' which corresponds to the energy difference between the singlet and the doublet states ($E_s-E_d$, see diagrams in Fig.~\ref{fig3}(g)). We assign positive values of $E_s-E_d$ to loop-like resonances (doublet ground state) and negative values to anticrossing-like resonances (singlet ground state). The data show that the lifting of Coulomb valleys and the singlet-doublet QPT occur within the same range of $V_S$. In other words, the signatures of odd-parity regimes are erased from transport data at similar dot parameters, both above and below the induced superconducting gap. These results are reproduced in another device, see supplemental materials. We note that the normal-state tunneling amplitude, a useful scaling parameter, cannot be extracted from these data in a way similar to Ref.\onlinecite{lee2017prb} because superconductivity cannot be suppressed due to the high critical temperature and field of NbTiN. Also, $V_{bias} = \unit[1]{mV}$ is be below the bulk gap of NbTiN, however, the gap is soft in this device meaning there is a finite single-particle density of states at all energies.

\begin{figure}[ht]
  \includegraphics[width=\columnwidth]{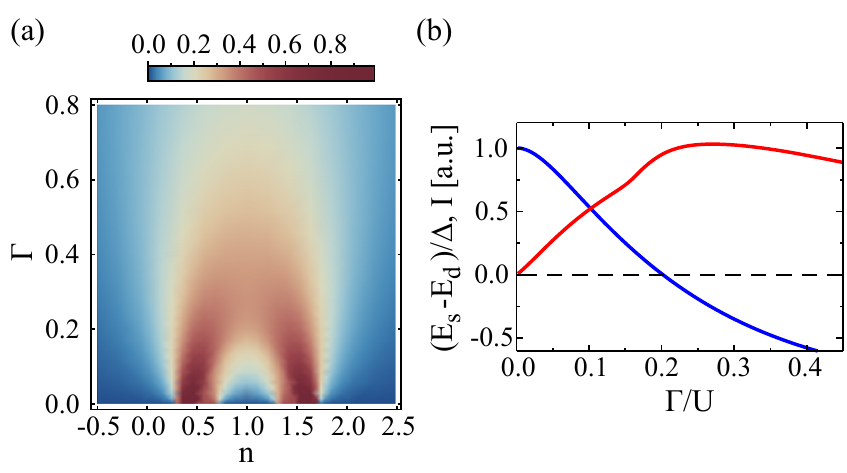}
  \caption{
(a) Simulated finite-bias current (in arbitrary units) computed as the integral over the quantum-dot spectral function as a function of the gate voltage (converted to units of charge, $n$) and the hybridization strength $\Gamma$.
(b) Theoretical singlet-doublet excitation energy splitting (blue line) and finite-bias current (red line) at half-filling, $n=1$, to be compared with Fig.~3(g).
 }
 \label{fig4}
\end{figure}

\begin{figure}[ht]
  \includegraphics[width=\columnwidth]{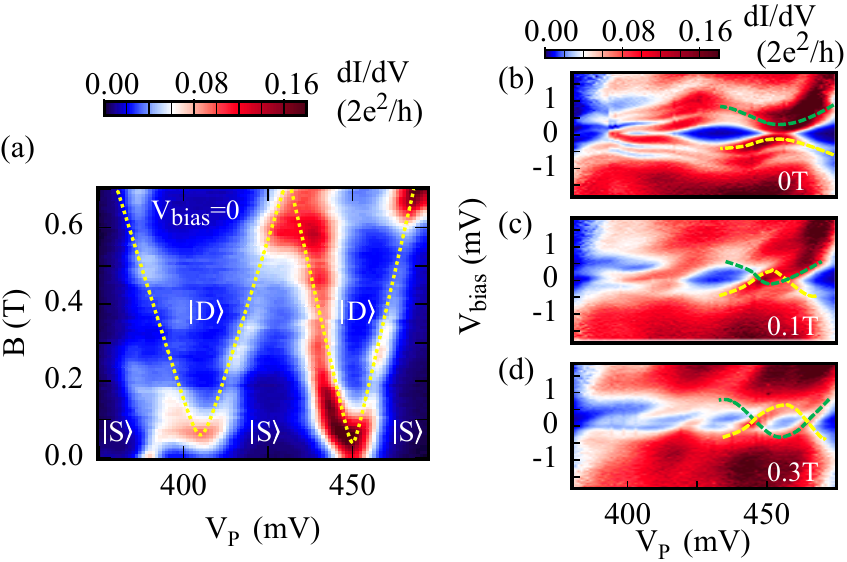}
  \caption{
(a) Zero-bias conductance for a range of magnetic field and 'p' gate settings. Ground states of the quantum dot are labeled $\ket{S}$ (singlet) and $\ket{D}$ (doublet). The yellow dashed traces are plots topological superconductivity condition scaled to match the high current contours. (b)-(d) Finite-bias spectroscopy for three values of the magnetic field across the singlet-doublet transition. Yellow and green traces are guides to the eye.
 }
 \label{fig5}
\end{figure}

In order to numerically analyze the observed experimental coincidence presented in Fig. \ref{fig3}(g), we describe the quantum dot using the single-impurity Anderson model:
\begin{equation}
H=\sum_\sigma \epsilon\, n_{\sigma} + U n_\uparrow n_\downarrow
+ \sum_{k\sigma} (t_k d^\dag_{\sigma} c_{k\sigma} + \text{H.c.})
+ H_\mathrm{lead},
\end{equation}
where $\epsilon$ is the quantum-dot level, $n_\sigma=d^\dag_\sigma
d_\sigma$ is the dot occupancy, $U$ is the electron-electron repulsion
parameter, $t_k$ are the tunneling amplitudes between the dot and the
superconducting lead. The coupling between the dot and the lead is
quantified by the hybridization strength $\Gamma=\pi |t_k|^2 \rho$,
where $\rho$ is the normal-state density of states in the electrode. The lead is
described by a Bardeen-Cooper-Schrieffer (BCS) Hamiltonian with the gap $\Delta$. In addition, we
take into account that the proximitized gap in the quantum dot is
soft. The complete information about the lead is contained in the 
hybridization function in the Nambu formalism. For a BCS Hamiltonian
it takes the form
\begin{equation}
\Delta(z) = \sum_{k} \frac{|t_k|^2}{z^2-\xi_k^2-\Delta^2}
\begin{bmatrix}
z+\xi_k & \Delta \\
\Delta & z-\xi_k
\end{bmatrix}.
\end{equation}
To describe a hard induced gap, one uses the argument $z=\omega+i0^+$. To
model a soft-gap situation, one instead uses $z=\omega+i\delta$ where
$\delta$ is the finite life-time of Bogoliubov quasiparticles: this
smoothens the BCS coherence peaks as well as produces a finite density
of states inside the gap.
We study the spectral and transport properties of this model using the
NRG method \cite{wilson1975,bulla2008,pillet_prb2013}, which consists of discretizing
the continuum of states in the leads, transforming the problem into a
tight-binding chain form, and iteratively diagonalizing the resulting
Hamiltonian. The Wilson chain coefficients that
correspond to the generalized BCS hybridization can be computed using
the artifact-less discretization scheme \cite{odesolv} generalized
to the matrix case \cite{liu2016}.
The spectral function is computed using the density-matrix NRG algorithm.
We used model parameters $U=1$, $\Delta=0.1$, systematically sweeping $\Gamma$ and $\epsilon$. The energy unit is half-bandwidth, $D=1$. The calculations were performed with
the NRG discretization parameter $\Lambda=2$ by averaging over $N_z=8$ interleaved
discretization grids.

To approximate the experimental finite-bias transport current within our model we compute an integral of the quantum dot spectral function in the spirit of the Meir-Wingreen formalism, neglecting any possible non-equilibrium effects, from $-2\Delta$ to $2\Delta$. These results are shown in Fig.~\ref{fig4}(a) as a function of the 'p' gate (converted to units of quantum dot charge $n$) and the contact barrier (expressed as hybridization strength $\Gamma$). The results can be interpreted based on the established knowledge about the transport through quantum dots with normal-state leads, taking into account that the gap $\Delta$ terminates the renormalization of the exchange coupling, thus it plays a similar role as the temperature $T$ in the normal-state case. Such $\Delta \leftrightarrow T$ mapping is possible, because the bias voltage is significantly larger than the gap and similar transport processes are relevant as in the normal state. Hence, in the weak coupling regime, where $T_K \ll \Delta$, the current is high only at the charge degeneracy points ($n=1/2$ and $n=3/2$) due to Coulomb blockade physics relevant in the $T_K \ll T$ regime. As the coupling increases, so that $T_K \sim \Delta$, the current increases in the valley center ($n=1$), as expected in the emerging Kondo regime; this is the range of $\Gamma/U \approx 0.2$ in Fig.~\ref{fig4}(a). In this regime, 1e-1e periodicity is no longer observed. For very open dot, the Kondo picture eventually breaks down due to strong charge fluctuations, for this reason the computed current decreases. The detailed behavior at $n=1$ as a function of $\Gamma$ is shown in Fig.~\ref{fig4}(b), where we furthermore show the energies of the sub-gap Andreev bound states, which are extracted directly from the NRG flow diagrams. We find that the singlet-doublet transition occurs at $\Gamma/U \sim 0.2$. For systems with $U \gg \Delta$, as is the case here, this transition is well known to be controlled by the ratio between the Kondo temperature and the superconducting gap, and occurs for $T_K \sim \Delta$. This point nearly coincides with the current maximum ($\Gamma/U \sim 0.2$ vs. $0.25$) and agrees at the qualitative level with the experimental results shown in Fig.~\ref{fig3}(b). This agreement is due to the fact that both phenomena occur at the point of maximal competition between the Kondo screening and superconducting pairing, thus they share common origin.
We observe some quantitative differences: in experiment, the current seems to reach a plateau at large $V_S$, while in simulation it decreases at higher $\Gamma$. This might be due to multiple occupied quantum dot levels, which are not included in the simplified theoretical model but are present in experiment.

We now discuss experimental data on the singlet-doublet QPT driven by Zeeman splitting, and its relevance to the studies of Majorana bound states (MBS). We position the 'p' and 's' gates in the strong coupled odd regime, where the ground state of the quantum dot is a spin-singlet at zero magnetic field (Fig. ~\ref{fig5}(a),(b)). Previous studies have shown that upon applying magnetic field a singlet-to-doublet QPT occurs. It is is marked by zero-bias peaks that appear when magnetic field shifts ABS to the middle of the gap \cite{LeeNatnano2014}. At the same time, zero-bias peaks are studied as signatures of MBS which accompany a  topological QPT. The topologically superconducting regime with MBS in infinite 1D nanowires is predicted for $E_Z > \sqrt[]{\mu^2+\Delta^2}$, where $E_Z$ is the Zeeman energy, $\mu$ is the chemical potential that corresponds to a gate voltage \cite{LutchynPRL2010, OregPRL2010}. The topological region within corresponds to an odd number of one-dimensional subbands crossing the Fermi level. 

In Fig.~\ref{fig5}(a) we show that the singlet-doublet QPC maps out a similar phase boundary to that expected for topological superconductivity. The data in Fig. \ref{fig5}(a) are obtained at zero bias, thus contours of high current correspond to zero-bias peaks. The ground states of the quantum dot are marked with $\ket{S}$ for singlet and $\ket{D}$ for doublet. Indeed we observe that the boundary between $\ket{S}$ and $\ket{D}$ regions has a shape similar to the topological phase boundary. Figures~\ref{fig5}(b)-(d) show the magnetic-field driven singlet-doublet QPT from anticrossing-like to loop-like resonances extended in the source-drain bias dimension. 

One key difference between the topological phase of an infinite nanowire and the magnetically-driven singlet-doublet QPT is that conductance inside the $\ket{D}$ region is low because the ABS resonances have shifted away from zero bias there, as indicated in Fig.~\ref{fig5}(d) where zero-bias peaks are strictly transient. In contrast, MBS resonances should remain at or near zero bias over an area in gate-vs-field map enclosed by the yellow dash-dot trace in Fig.~\ref{fig5}(a). Note that here the magnetic field is applied at 30 degrees with respect to the nanowire, so no MBS are expected because the field has a significant component parallel to the spin-orbit field. However, since a small quantum dot is explicitly defined here by gates, we do not expect MBS to appear. We furthermore remark that the zero-bias conductance is not precisely zero due to resonance broadening and soft gap. Nevertheless, Fig.~\ref{fig5}(a) adds to the list of known experimental similarities between ABS and MBS.

\textit{Acknowledgements.} We thank D. Pekker, P. Yu, A. Zarassi for discussions. S.M.F. is supported by NSF DMR-1743972, NSF PIRE-1743717, ONR and ARO. R.\v{Z}. acknowledges support from the Slovenian Research Agency (ARRS) under Grants No. P1-0044 and J1-7259.

\bibliographystyle{apsrev4-1}
\bibliography{Ref.bib}

\begin{thebibliography}{40}%
\makeatletter
\providecommand \@ifxundefined [1]{%
 \@ifx{#1\undefined}
}%
\providecommand \@ifnum [1]{%
 \ifnum #1\expandafter \@firstoftwo
 \else \expandafter \@secondoftwo
 \fi
}%
\providecommand \@ifx [1]{%
 \ifx #1\expandafter \@firstoftwo
 \else \expandafter \@secondoftwo
 \fi
}%
\providecommand \natexlab [1]{#1}%
\providecommand \enquote  [1]{``#1''}%
\providecommand \bibnamefont  [1]{#1}%
\providecommand \bibfnamefont [1]{#1}%
\providecommand \citenamefont [1]{#1}%
\providecommand \href@noop [0]{\@secondoftwo}%
\providecommand \href [0]{\begingroup \@sanitize@url \@href}%
\providecommand \@href[1]{\@@startlink{#1}\@@href}%
\providecommand \@@href[1]{\endgroup#1\@@endlink}%
\providecommand \@sanitize@url [0]{\catcode `\\12\catcode `\$12\catcode
  `\&12\catcode `\#12\catcode `\^12\catcode `\_12\catcode `\%12\relax}%
\providecommand \@@startlink[1]{}%
\providecommand \@@endlink[0]{}%
\providecommand \url  [0]{\begingroup\@sanitize@url \@url }%
\providecommand \@url [1]{\endgroup\@href {#1}{\urlprefix }}%
\providecommand \urlprefix  [0]{URL }%
\providecommand \Eprint [0]{\href }%
\providecommand \doibase [0]{http://dx.doi.org/}%
\providecommand \selectlanguage [0]{\@gobble}%
\providecommand \bibinfo  [0]{\@secondoftwo}%
\providecommand \bibfield  [0]{\@secondoftwo}%
\providecommand \translation [1]{[#1]}%
\providecommand \BibitemOpen [0]{}%
\providecommand \bibitemStop [0]{}%
\providecommand \bibitemNoStop [0]{.\EOS\space}%
\providecommand \EOS [0]{\spacefactor3000\relax}%
\providecommand \BibitemShut  [1]{\csname bibitem#1\endcsname}%
\let\auto@bib@innerbib\@empty
\bibitem [{\citenamefont {Das~Sarma}\ \emph {et~al.}(2015)\citenamefont
  {Das~Sarma}, \citenamefont {Freedman},\ and\ \citenamefont
  {Nayak}}]{sarma2015majorana}%
  \BibitemOpen
  \bibfield  {author} {\bibinfo {author} {\bibfnamefont {S.}~\bibnamefont
  {Das~Sarma}}, \bibinfo {author} {\bibfnamefont {M.}~\bibnamefont {Freedman}},
  \ and\ \bibinfo {author} {\bibfnamefont {C.}~\bibnamefont {Nayak}},\
  }\href@noop {} {\bibfield  {journal} {\bibinfo  {journal} {npj Quantum
  Information}\ }\textbf {\bibinfo {volume} {1}},\ \bibinfo {pages} {15001}
  (\bibinfo {year} {2015})}\BibitemShut {NoStop}%
\bibitem [{\citenamefont {Tuominen}\ \emph {et~al.}(1992)\citenamefont
  {Tuominen}, \citenamefont {Hergenrother}, \citenamefont {Tighe},\ and\
  \citenamefont {Tinkham}}]{TuominenPRL92}%
  \BibitemOpen
  \bibfield  {author} {\bibinfo {author} {\bibfnamefont {M.~T.}\ \bibnamefont
  {Tuominen}}, \bibinfo {author} {\bibfnamefont {J.~M.}\ \bibnamefont
  {Hergenrother}}, \bibinfo {author} {\bibfnamefont {T.~S.}\ \bibnamefont
  {Tighe}}, \ and\ \bibinfo {author} {\bibfnamefont {M.}~\bibnamefont
  {Tinkham}},\ }\href {\doibase 10.1103/PhysRevLett.69.1997} {\bibfield
  {journal} {\bibinfo  {journal} {Phys. Rev. Lett.}\ }\textbf {\bibinfo
  {volume} {69}},\ \bibinfo {pages} {1997} (\bibinfo {year}
  {1992})}\BibitemShut {NoStop}%
\bibitem [{\citenamefont {Eiles}\ \emph {et~al.}(1993)\citenamefont {Eiles},
  \citenamefont {Martinis},\ and\ \citenamefont {Devoret}}]{eilesPRL93}%
  \BibitemOpen
  \bibfield  {author} {\bibinfo {author} {\bibfnamefont {T.~M.}\ \bibnamefont
  {Eiles}}, \bibinfo {author} {\bibfnamefont {J.~M.}\ \bibnamefont {Martinis}},
  \ and\ \bibinfo {author} {\bibfnamefont {M.~H.}\ \bibnamefont {Devoret}},\
  }\href {\doibase 10.1103/PhysRevLett.70.1862} {\bibfield  {journal} {\bibinfo
   {journal} {Phys. Rev. Lett.}\ }\textbf {\bibinfo {volume} {70}},\ \bibinfo
  {pages} {1862} (\bibinfo {year} {1993})}\BibitemShut {NoStop}%
\bibitem [{\citenamefont {Lafarge}\ \emph {et~al.}(1993)\citenamefont
  {Lafarge}, \citenamefont {Joyez}, \citenamefont {Esteve}, \citenamefont
  {Urbina},\ and\ \citenamefont {Devoret}}]{LafargePRL93}%
  \BibitemOpen
  \bibfield  {author} {\bibinfo {author} {\bibfnamefont {P.}~\bibnamefont
  {Lafarge}}, \bibinfo {author} {\bibfnamefont {P.}~\bibnamefont {Joyez}},
  \bibinfo {author} {\bibfnamefont {D.}~\bibnamefont {Esteve}}, \bibinfo
  {author} {\bibfnamefont {C.}~\bibnamefont {Urbina}}, \ and\ \bibinfo {author}
  {\bibfnamefont {M.~H.}\ \bibnamefont {Devoret}},\ }\href {\doibase
  10.1103/PhysRevLett.70.994} {\bibfield  {journal} {\bibinfo  {journal} {Phys.
  Rev. Lett.}\ }\textbf {\bibinfo {volume} {70}},\ \bibinfo {pages} {994}
  (\bibinfo {year} {1993})}\BibitemShut {NoStop}%
\bibitem [{\citenamefont {Albrecht}\ \emph {et~al.}(2016)\citenamefont
  {Albrecht}, \citenamefont {Higginbotham}, \citenamefont {Madsen},
  \citenamefont {Kuemmeth}, \citenamefont {Jespersen}, \citenamefont
  {Nyg\r{a}rd}, \citenamefont {Krogstrup},\ and\ \citenamefont
  {Marcus}}]{AlbrechtNature2016}%
  \BibitemOpen
  \bibfield  {author} {\bibinfo {author} {\bibfnamefont {S.~M.}\ \bibnamefont
  {Albrecht}}, \bibinfo {author} {\bibfnamefont {A.~P.}\ \bibnamefont
  {Higginbotham}}, \bibinfo {author} {\bibfnamefont {M.}~\bibnamefont
  {Madsen}}, \bibinfo {author} {\bibfnamefont {F.}~\bibnamefont {Kuemmeth}},
  \bibinfo {author} {\bibfnamefont {T.~S.}\ \bibnamefont {Jespersen}}, \bibinfo
  {author} {\bibfnamefont {J.}~\bibnamefont {Nyg\r{a}rd}}, \bibinfo {author}
  {\bibfnamefont {P.}~\bibnamefont {Krogstrup}}, \ and\ \bibinfo {author}
  {\bibfnamefont {C.~M.}\ \bibnamefont {Marcus}},\ }\href@noop {} {\bibfield
  {journal} {\bibinfo  {journal} {Nature}\ }\textbf {\bibinfo {volume} {531}},\
  \bibinfo {pages} {206} (\bibinfo {year} {2016})}\BibitemShut {NoStop}%
\bibitem [{\citenamefont {Janvier}\ \emph {et~al.}(2015)\citenamefont
  {Janvier}, \citenamefont {Tosi}, \citenamefont {Bretheau}, \citenamefont
  {Girit}, \citenamefont {Stern}, \citenamefont {Bertet}, \citenamefont
  {Joyez}, \citenamefont {Vion}, \citenamefont {Esteve}, \citenamefont
  {Goffman} \emph {et~al.}}]{janvierscience15}%
  \BibitemOpen
  \bibfield  {author} {\bibinfo {author} {\bibfnamefont {C.}~\bibnamefont
  {Janvier}}, \bibinfo {author} {\bibfnamefont {L.}~\bibnamefont {Tosi}},
  \bibinfo {author} {\bibfnamefont {L.}~\bibnamefont {Bretheau}}, \bibinfo
  {author} {\bibfnamefont {{\c{C}}.}~\bibnamefont {Girit}}, \bibinfo {author}
  {\bibfnamefont {M.}~\bibnamefont {Stern}}, \bibinfo {author} {\bibfnamefont
  {P.}~\bibnamefont {Bertet}}, \bibinfo {author} {\bibfnamefont
  {P.}~\bibnamefont {Joyez}}, \bibinfo {author} {\bibfnamefont
  {D.}~\bibnamefont {Vion}}, \bibinfo {author} {\bibfnamefont {D.}~\bibnamefont
  {Esteve}}, \bibinfo {author} {\bibfnamefont {M.}~\bibnamefont {Goffman}},
  \emph {et~al.},\ }\href@noop {} {\bibfield  {journal} {\bibinfo  {journal}
  {Science}\ }\textbf {\bibinfo {volume} {349}},\ \bibinfo {pages} {1199}
  (\bibinfo {year} {2015})}\BibitemShut {NoStop}%
\bibitem [{\citenamefont {Hays}\ \emph {et~al.}(2018)\citenamefont {Hays},
  \citenamefont {de~Lange}, \citenamefont {Serniak}, \citenamefont {van
  Woerkom}, \citenamefont {Bouman}, \citenamefont {Krogstrup}, \citenamefont
  {Nyg\aa{}rd}, \citenamefont {Geresdi},\ and\ \citenamefont
  {Devoret}}]{haysprl18}%
  \BibitemOpen
  \bibfield  {author} {\bibinfo {author} {\bibfnamefont {M.}~\bibnamefont
  {Hays}}, \bibinfo {author} {\bibfnamefont {G.}~\bibnamefont {de~Lange}},
  \bibinfo {author} {\bibfnamefont {K.}~\bibnamefont {Serniak}}, \bibinfo
  {author} {\bibfnamefont {D.~J.}\ \bibnamefont {van Woerkom}}, \bibinfo
  {author} {\bibfnamefont {D.}~\bibnamefont {Bouman}}, \bibinfo {author}
  {\bibfnamefont {P.}~\bibnamefont {Krogstrup}}, \bibinfo {author}
  {\bibfnamefont {J.}~\bibnamefont {Nyg\aa{}rd}}, \bibinfo {author}
  {\bibfnamefont {A.}~\bibnamefont {Geresdi}}, \ and\ \bibinfo {author}
  {\bibfnamefont {M.~H.}\ \bibnamefont {Devoret}},\ }\href {\doibase
  10.1103/PhysRevLett.121.047001} {\bibfield  {journal} {\bibinfo  {journal}
  {Phys. Rev. Lett.}\ }\textbf {\bibinfo {volume} {121}},\ \bibinfo {pages}
  {047001} (\bibinfo {year} {2018})}\BibitemShut {NoStop}%
\bibitem [{\citenamefont {Matsuura}(1977)}]{matsuura1977}%
  \BibitemOpen
  \bibfield  {author} {\bibinfo {author} {\bibfnamefont {T.}~\bibnamefont
  {Matsuura}},\ }\href@noop {} {\bibfield  {journal} {\bibinfo  {journal}
  {Prog. Theor. Phys.}\ }\textbf {\bibinfo {volume} {57}},\ \bibinfo {pages}
  {1823} (\bibinfo {year} {1977})}\BibitemShut {NoStop}%
\bibitem [{\citenamefont {Satori}\ \emph {et~al.}(1992)\citenamefont {Satori},
  \citenamefont {Shiba}, \citenamefont {Sakai},\ and\ \citenamefont
  {Shimizu}}]{satori1992}%
  \BibitemOpen
  \bibfield  {author} {\bibinfo {author} {\bibfnamefont {K.}~\bibnamefont
  {Satori}}, \bibinfo {author} {\bibfnamefont {H.}~\bibnamefont {Shiba}},
  \bibinfo {author} {\bibfnamefont {O.}~\bibnamefont {Sakai}}, \ and\ \bibinfo
  {author} {\bibfnamefont {Y.}~\bibnamefont {Shimizu}},\ }\href@noop {}
  {\bibfield  {journal} {\bibinfo  {journal} {J. Phys. Soc. Japan}\ }\textbf
  {\bibinfo {volume} {61}},\ \bibinfo {pages} {3239} (\bibinfo {year}
  {1992})}\BibitemShut {NoStop}%
\bibitem [{\citenamefont {Yoshioka}\ and\ \citenamefont
  {Ohashi}(2000)}]{yoshioka2000}%
  \BibitemOpen
  \bibfield  {author} {\bibinfo {author} {\bibfnamefont {T.}~\bibnamefont
  {Yoshioka}}\ and\ \bibinfo {author} {\bibfnamefont {Y.}~\bibnamefont
  {Ohashi}},\ }\href@noop {} {\bibfield  {journal} {\bibinfo  {journal} {J.
  Phys. Soc. Japan}\ }\textbf {\bibinfo {volume} {69}},\ \bibinfo {pages}
  {1812} (\bibinfo {year} {2000})}\BibitemShut {NoStop}%
\bibitem [{\citenamefont {Choi}\ \emph {et~al.}(2004)\citenamefont {Choi},
  \citenamefont {Lee}, \citenamefont {Kang},\ and\ \citenamefont
  {Belzig}}]{choi2004josephson}%
  \BibitemOpen
  \bibfield  {author} {\bibinfo {author} {\bibfnamefont {M.-S.}\ \bibnamefont
  {Choi}}, \bibinfo {author} {\bibfnamefont {M.}~\bibnamefont {Lee}}, \bibinfo
  {author} {\bibfnamefont {K.}~\bibnamefont {Kang}}, \ and\ \bibinfo {author}
  {\bibfnamefont {W.}~\bibnamefont {Belzig}},\ }\href@noop {} {\bibfield
  {journal} {\bibinfo  {journal} {Phys. Rev. B}\ }\textbf {\bibinfo {volume}
  {70}},\ \bibinfo {pages} {020502} (\bibinfo {year} {2004})}\BibitemShut
  {NoStop}%
\bibitem [{\citenamefont {Oguri}\ and\ \citenamefont
  {Tanaka}(2004)}]{oguri2004josephson}%
  \BibitemOpen
  \bibfield  {author} {\bibinfo {author} {\bibfnamefont {A.}~\bibnamefont
  {Oguri}}\ and\ \bibinfo {author} {\bibfnamefont {Y.}~\bibnamefont {Tanaka}},\
  }\href@noop {} {\bibfield  {journal} {\bibinfo  {journal} {Journal of the
  Physical Society of Japan}\ }\textbf {\bibinfo {volume} {73}},\ \bibinfo
  {pages} {2494} (\bibinfo {year} {2004})}\BibitemShut {NoStop}%
\bibitem [{\citenamefont {Bauer}\ \emph {et~al.}(2007)\citenamefont {Bauer},
  \citenamefont {Oguri},\ and\ \citenamefont {Hewson}}]{bauer2007}%
  \BibitemOpen
  \bibfield  {author} {\bibinfo {author} {\bibfnamefont {J.}~\bibnamefont
  {Bauer}}, \bibinfo {author} {\bibfnamefont {A.}~\bibnamefont {Oguri}}, \ and\
  \bibinfo {author} {\bibfnamefont {A.~C.}\ \bibnamefont {Hewson}},\
  }\href@noop {} {\bibfield  {journal} {\bibinfo  {journal} {J. Phys.: Condens.
  Matter}\ }\textbf {\bibinfo {volume} {19}},\ \bibinfo {pages} {486211}
  (\bibinfo {year} {2007})}\BibitemShut {NoStop}%
\bibitem [{\citenamefont {Karrasch}\ \emph {et~al.}(2008)\citenamefont
  {Karrasch}, \citenamefont {Oguri},\ and\ \citenamefont
  {Meden}}]{karrasch2008}%
  \BibitemOpen
  \bibfield  {author} {\bibinfo {author} {\bibfnamefont {C.}~\bibnamefont
  {Karrasch}}, \bibinfo {author} {\bibfnamefont {A.}~\bibnamefont {Oguri}}, \
  and\ \bibinfo {author} {\bibfnamefont {V.}~\bibnamefont {Meden}},\
  }\href@noop {} {\bibfield  {journal} {\bibinfo  {journal} {Phys. Rev. B}\
  }\textbf {\bibinfo {volume} {77}},\ \bibinfo {pages} {024517} (\bibinfo
  {year} {2008})}\BibitemShut {NoStop}%
\bibitem [{\citenamefont {Eichler}\ \emph {et~al.}(2007)\citenamefont
  {Eichler}, \citenamefont {Weiss}, \citenamefont {Oberholzer}, \citenamefont
  {Sch\"onenberger}, \citenamefont {Levy~Yeyati}, \citenamefont {Cuevas},\ and\
  \citenamefont {Mart\'{\i}n-Rodero}}]{EichlerPRL07}%
  \BibitemOpen
  \bibfield  {author} {\bibinfo {author} {\bibfnamefont {A.}~\bibnamefont
  {Eichler}}, \bibinfo {author} {\bibfnamefont {M.}~\bibnamefont {Weiss}},
  \bibinfo {author} {\bibfnamefont {S.}~\bibnamefont {Oberholzer}}, \bibinfo
  {author} {\bibfnamefont {C.}~\bibnamefont {Sch\"onenberger}}, \bibinfo
  {author} {\bibfnamefont {A.}~\bibnamefont {Levy~Yeyati}}, \bibinfo {author}
  {\bibfnamefont {J.~C.}\ \bibnamefont {Cuevas}}, \ and\ \bibinfo {author}
  {\bibfnamefont {A.}~\bibnamefont {Mart\'{\i}n-Rodero}},\ }\href {\doibase
  10.1103/PhysRevLett.99.126602} {\bibfield  {journal} {\bibinfo  {journal}
  {Phys. Rev. Lett.}\ }\textbf {\bibinfo {volume} {99}},\ \bibinfo {pages}
  {126602} (\bibinfo {year} {2007})}\BibitemShut {NoStop}%
\bibitem [{\citenamefont {Sand-Jespersen}\ \emph {et~al.}(2007)\citenamefont
  {Sand-Jespersen}, \citenamefont {Paaske}, \citenamefont {Andersen},
  \citenamefont {Grove-Rasmussen}, \citenamefont {J\o{}rgensen}, \citenamefont
  {Aagesen}, \citenamefont {S\o{}rensen}, \citenamefont {Lindelof},
  \citenamefont {Flensberg},\ and\ \citenamefont
  {Nyg\aa{}rd}}]{SandJespersenPRL07}%
  \BibitemOpen
  \bibfield  {author} {\bibinfo {author} {\bibfnamefont {T.}~\bibnamefont
  {Sand-Jespersen}}, \bibinfo {author} {\bibfnamefont {J.}~\bibnamefont
  {Paaske}}, \bibinfo {author} {\bibfnamefont {B.~M.}\ \bibnamefont
  {Andersen}}, \bibinfo {author} {\bibfnamefont {K.}~\bibnamefont
  {Grove-Rasmussen}}, \bibinfo {author} {\bibfnamefont {H.~I.}\ \bibnamefont
  {J\o{}rgensen}}, \bibinfo {author} {\bibfnamefont {M.}~\bibnamefont
  {Aagesen}}, \bibinfo {author} {\bibfnamefont {C.~B.}\ \bibnamefont
  {S\o{}rensen}}, \bibinfo {author} {\bibfnamefont {P.~E.}\ \bibnamefont
  {Lindelof}}, \bibinfo {author} {\bibfnamefont {K.}~\bibnamefont {Flensberg}},
  \ and\ \bibinfo {author} {\bibfnamefont {J.}~\bibnamefont {Nyg\aa{}rd}},\
  }\href {\doibase 10.1103/PhysRevLett.99.126603} {\bibfield  {journal}
  {\bibinfo  {journal} {Phys. Rev. Lett.}\ }\textbf {\bibinfo {volume} {99}},\
  \bibinfo {pages} {126603} (\bibinfo {year} {2007})}\BibitemShut {NoStop}%
\bibitem [{\citenamefont {Buizert}\ \emph {et~al.}(2007)\citenamefont
  {Buizert}, \citenamefont {Oiwa}, \citenamefont {Shibata}, \citenamefont
  {Hirakawa},\ and\ \citenamefont {Tarucha}}]{BuizertPRL07}%
  \BibitemOpen
  \bibfield  {author} {\bibinfo {author} {\bibfnamefont {C.}~\bibnamefont
  {Buizert}}, \bibinfo {author} {\bibfnamefont {A.}~\bibnamefont {Oiwa}},
  \bibinfo {author} {\bibfnamefont {K.}~\bibnamefont {Shibata}}, \bibinfo
  {author} {\bibfnamefont {K.}~\bibnamefont {Hirakawa}}, \ and\ \bibinfo
  {author} {\bibfnamefont {S.}~\bibnamefont {Tarucha}},\ }\href {\doibase
  10.1103/PhysRevLett.99.136806} {\bibfield  {journal} {\bibinfo  {journal}
  {Phys. Rev. Lett.}\ }\textbf {\bibinfo {volume} {99}},\ \bibinfo {pages}
  {136806} (\bibinfo {year} {2007})}\BibitemShut {NoStop}%
\bibitem [{\citenamefont {Grove-Rasmussen}\ \emph {et~al.}(2009)\citenamefont
  {Grove-Rasmussen}, \citenamefont {J\o{}rgensen}, \citenamefont {Andersen},
  \citenamefont {Paaske}, \citenamefont {Jespersen}, \citenamefont
  {Nyg\aa{}rd}, \citenamefont {Flensberg},\ and\ \citenamefont
  {Lindelof}}]{GroveRasmussenPRB09}%
  \BibitemOpen
  \bibfield  {author} {\bibinfo {author} {\bibfnamefont {K.}~\bibnamefont
  {Grove-Rasmussen}}, \bibinfo {author} {\bibfnamefont {H.~I.}\ \bibnamefont
  {J\o{}rgensen}}, \bibinfo {author} {\bibfnamefont {B.~M.}\ \bibnamefont
  {Andersen}}, \bibinfo {author} {\bibfnamefont {J.}~\bibnamefont {Paaske}},
  \bibinfo {author} {\bibfnamefont {T.~S.}\ \bibnamefont {Jespersen}}, \bibinfo
  {author} {\bibfnamefont {J.}~\bibnamefont {Nyg\aa{}rd}}, \bibinfo {author}
  {\bibfnamefont {K.}~\bibnamefont {Flensberg}}, \ and\ \bibinfo {author}
  {\bibfnamefont {P.~E.}\ \bibnamefont {Lindelof}},\ }\href {\doibase
  10.1103/PhysRevB.79.134518} {\bibfield  {journal} {\bibinfo  {journal} {Phys.
  Rev. B}\ }\textbf {\bibinfo {volume} {79}},\ \bibinfo {pages} {134518}
  (\bibinfo {year} {2009})}\BibitemShut {NoStop}%
\bibitem [{\citenamefont {Deacon}\ \emph {et~al.}(2010)\citenamefont {Deacon},
  \citenamefont {Tanaka}, \citenamefont {Oiwa}, \citenamefont {Sakano},
  \citenamefont {Yoshida}, \citenamefont {Shibata}, \citenamefont {Hirakawa},\
  and\ \citenamefont {Tarucha}}]{DeaconPRL10}%
  \BibitemOpen
  \bibfield  {author} {\bibinfo {author} {\bibfnamefont {R.~S.}\ \bibnamefont
  {Deacon}}, \bibinfo {author} {\bibfnamefont {Y.}~\bibnamefont {Tanaka}},
  \bibinfo {author} {\bibfnamefont {A.}~\bibnamefont {Oiwa}}, \bibinfo {author}
  {\bibfnamefont {R.}~\bibnamefont {Sakano}}, \bibinfo {author} {\bibfnamefont
  {K.}~\bibnamefont {Yoshida}}, \bibinfo {author} {\bibfnamefont
  {K.}~\bibnamefont {Shibata}}, \bibinfo {author} {\bibfnamefont
  {K.}~\bibnamefont {Hirakawa}}, \ and\ \bibinfo {author} {\bibfnamefont
  {S.}~\bibnamefont {Tarucha}},\ }\href {\doibase
  10.1103/PhysRevLett.104.076805} {\bibfield  {journal} {\bibinfo  {journal}
  {Phys. Rev. Lett.}\ }\textbf {\bibinfo {volume} {104}},\ \bibinfo {pages}
  {076805} (\bibinfo {year} {2010})}\BibitemShut {NoStop}%
\bibitem [{\citenamefont {Kanai}\ \emph {et~al.}(2010)\citenamefont {Kanai},
  \citenamefont {Deacon}, \citenamefont {Oiwa}, \citenamefont {Yoshida},
  \citenamefont {Shibata}, \citenamefont {Hirakawa},\ and\ \citenamefont
  {Tarucha}}]{KanaiPRB10}%
  \BibitemOpen
  \bibfield  {author} {\bibinfo {author} {\bibfnamefont {Y.}~\bibnamefont
  {Kanai}}, \bibinfo {author} {\bibfnamefont {R.~S.}\ \bibnamefont {Deacon}},
  \bibinfo {author} {\bibfnamefont {A.}~\bibnamefont {Oiwa}}, \bibinfo {author}
  {\bibfnamefont {K.}~\bibnamefont {Yoshida}}, \bibinfo {author} {\bibfnamefont
  {K.}~\bibnamefont {Shibata}}, \bibinfo {author} {\bibfnamefont
  {K.}~\bibnamefont {Hirakawa}}, \ and\ \bibinfo {author} {\bibfnamefont
  {S.}~\bibnamefont {Tarucha}},\ }\href {\doibase 10.1103/PhysRevB.82.054512}
  {\bibfield  {journal} {\bibinfo  {journal} {Phys. Rev. B}\ }\textbf {\bibinfo
  {volume} {82}},\ \bibinfo {pages} {054512} (\bibinfo {year}
  {2010})}\BibitemShut {NoStop}%
\bibitem [{\citenamefont {Pillet}\ \emph {et~al.}(2010)\citenamefont {Pillet},
  \citenamefont {Quay}, \citenamefont {Morfin}, \citenamefont {Bena},
  \citenamefont {Levy~Yeyati},\ and\ \citenamefont {Joyez}}]{first_ABS}%
  \BibitemOpen
  \bibfield  {author} {\bibinfo {author} {\bibfnamefont {J.-D.}\ \bibnamefont
  {Pillet}}, \bibinfo {author} {\bibfnamefont {C.~H.~L.}\ \bibnamefont {Quay}},
  \bibinfo {author} {\bibfnamefont {P.}~\bibnamefont {Morfin}}, \bibinfo
  {author} {\bibfnamefont {C.}~\bibnamefont {Bena}}, \bibinfo {author}
  {\bibfnamefont {A.}~\bibnamefont {Levy~Yeyati}}, \ and\ \bibinfo {author}
  {\bibfnamefont {P.}~\bibnamefont {Joyez}},\ }\href@noop {} {\bibfield
  {journal} {\bibinfo  {journal} {Nature Phys.}\ }\textbf {\bibinfo {volume}
  {6}},\ \bibinfo {pages} {965} (\bibinfo {year} {2010})}\BibitemShut {NoStop}%
\bibitem [{\citenamefont {Maurand}\ \emph {et~al.}(2012)\citenamefont
  {Maurand}, \citenamefont {Meng}, \citenamefont {Bonet}, \citenamefont
  {Florens}, \citenamefont {Marty},\ and\ \citenamefont
  {Wernsdorfer}}]{MaurandPRX12}%
  \BibitemOpen
  \bibfield  {author} {\bibinfo {author} {\bibfnamefont {R.}~\bibnamefont
  {Maurand}}, \bibinfo {author} {\bibfnamefont {T.}~\bibnamefont {Meng}},
  \bibinfo {author} {\bibfnamefont {E.}~\bibnamefont {Bonet}}, \bibinfo
  {author} {\bibfnamefont {S.}~\bibnamefont {Florens}}, \bibinfo {author}
  {\bibfnamefont {L.}~\bibnamefont {Marty}}, \ and\ \bibinfo {author}
  {\bibfnamefont {W.}~\bibnamefont {Wernsdorfer}},\ }\href {\doibase
  10.1103/PhysRevX.2.011009} {\bibfield  {journal} {\bibinfo  {journal} {Phys.
  Rev. X}\ }\textbf {\bibinfo {volume} {2}},\ \bibinfo {pages} {011009}
  (\bibinfo {year} {2012})}\BibitemShut {NoStop}%
\bibitem [{\citenamefont {Chang}\ \emph {et~al.}(2013)\citenamefont {Chang},
  \citenamefont {Manucharyan}, \citenamefont {Jespersen}, \citenamefont
  {Nyg\aa{}rd},\ and\ \citenamefont {Marcus}}]{ChangPRL13}%
  \BibitemOpen
  \bibfield  {author} {\bibinfo {author} {\bibfnamefont {W.}~\bibnamefont
  {Chang}}, \bibinfo {author} {\bibfnamefont {V.~E.}\ \bibnamefont
  {Manucharyan}}, \bibinfo {author} {\bibfnamefont {T.~S.}\ \bibnamefont
  {Jespersen}}, \bibinfo {author} {\bibfnamefont {J.}~\bibnamefont
  {Nyg\aa{}rd}}, \ and\ \bibinfo {author} {\bibfnamefont {C.~M.}\ \bibnamefont
  {Marcus}},\ }\href {\doibase 10.1103/PhysRevLett.110.217005} {\bibfield
  {journal} {\bibinfo  {journal} {Phys. Rev. Lett.}\ }\textbf {\bibinfo
  {volume} {110}},\ \bibinfo {pages} {217005} (\bibinfo {year}
  {2013})}\BibitemShut {NoStop}%
\bibitem [{\citenamefont {Kumar}\ \emph {et~al.}(2014)\citenamefont {Kumar},
  \citenamefont {Gaim}, \citenamefont {Steininger}, \citenamefont {Yeyati},
  \citenamefont {Mart\'{\i}n-Rodero}, \citenamefont {H\"uttel},\ and\
  \citenamefont {Strunk}}]{KumarPRB14}%
  \BibitemOpen
  \bibfield  {author} {\bibinfo {author} {\bibfnamefont {A.}~\bibnamefont
  {Kumar}}, \bibinfo {author} {\bibfnamefont {M.}~\bibnamefont {Gaim}},
  \bibinfo {author} {\bibfnamefont {D.}~\bibnamefont {Steininger}}, \bibinfo
  {author} {\bibfnamefont {A.~L.}\ \bibnamefont {Yeyati}}, \bibinfo {author}
  {\bibfnamefont {A.}~\bibnamefont {Mart\'{\i}n-Rodero}}, \bibinfo {author}
  {\bibfnamefont {A.~K.}\ \bibnamefont {H\"uttel}}, \ and\ \bibinfo {author}
  {\bibfnamefont {C.}~\bibnamefont {Strunk}},\ }\href {\doibase
  10.1103/PhysRevB.89.075428} {\bibfield  {journal} {\bibinfo  {journal} {Phys.
  Rev. B}\ }\textbf {\bibinfo {volume} {89}},\ \bibinfo {pages} {075428}
  (\bibinfo {year} {2014})}\BibitemShut {NoStop}%
\bibitem [{\citenamefont {Lee}\ \emph {et~al.}(2014)\citenamefont {Lee},
  \citenamefont {Jiang}, \citenamefont {Houzet}, \citenamefont {Aguado},
  \citenamefont {Lieber},\ and\ \citenamefont
  {De~Franceschi}}]{LeeNatnano2014}%
  \BibitemOpen
  \bibfield  {author} {\bibinfo {author} {\bibfnamefont {E.~J.~H.}\
  \bibnamefont {Lee}}, \bibinfo {author} {\bibfnamefont {X.}~\bibnamefont
  {Jiang}}, \bibinfo {author} {\bibfnamefont {M.}~\bibnamefont {Houzet}},
  \bibinfo {author} {\bibfnamefont {R.}~\bibnamefont {Aguado}}, \bibinfo
  {author} {\bibfnamefont {C.~M.}\ \bibnamefont {Lieber}}, \ and\ \bibinfo
  {author} {\bibfnamefont {S.}~\bibnamefont {De~Franceschi}},\ }\href@noop {}
  {\bibfield  {journal} {\bibinfo  {journal} {Nature nanotechnology}\ }\textbf
  {\bibinfo {volume} {9}},\ \bibinfo {pages} {79} (\bibinfo {year}
  {2014})}\BibitemShut {NoStop}%
\bibitem [{\citenamefont {Jellinggaard}\ \emph {et~al.}(2016)\citenamefont
  {Jellinggaard}, \citenamefont {Grove-Rasmussen}, \citenamefont {Madsen},\
  and\ \citenamefont {Nyg\aa{}rd}}]{JellinggaardPRB16}%
  \BibitemOpen
  \bibfield  {author} {\bibinfo {author} {\bibfnamefont {A.}~\bibnamefont
  {Jellinggaard}}, \bibinfo {author} {\bibfnamefont {K.}~\bibnamefont
  {Grove-Rasmussen}}, \bibinfo {author} {\bibfnamefont {M.~H.}\ \bibnamefont
  {Madsen}}, \ and\ \bibinfo {author} {\bibfnamefont {J.}~\bibnamefont
  {Nyg\aa{}rd}},\ }\href {\doibase 10.1103/PhysRevB.94.064520} {\bibfield
  {journal} {\bibinfo  {journal} {Phys. Rev. B}\ }\textbf {\bibinfo {volume}
  {94}},\ \bibinfo {pages} {064520} (\bibinfo {year} {2016})}\BibitemShut
  {NoStop}%
\bibitem [{\citenamefont {Lee}\ \emph {et~al.}(2017)\citenamefont {Lee},
  \citenamefont {Jiang}, \citenamefont {\ifmmode~\check{Z}\else
  \v{Z}\fi{}itko}, \citenamefont {Aguado}, \citenamefont {Lieber},\ and\
  \citenamefont {De~Franceschi}}]{lee2017prb}%
  \BibitemOpen
  \bibfield  {author} {\bibinfo {author} {\bibfnamefont {E.~J.~H.}\
  \bibnamefont {Lee}}, \bibinfo {author} {\bibfnamefont {X.}~\bibnamefont
  {Jiang}}, \bibinfo {author} {\bibfnamefont {R.}~\bibnamefont
  {\ifmmode~\check{Z}\else \v{Z}\fi{}itko}}, \bibinfo {author} {\bibfnamefont
  {R.}~\bibnamefont {Aguado}}, \bibinfo {author} {\bibfnamefont {C.~M.}\
  \bibnamefont {Lieber}}, \ and\ \bibinfo {author} {\bibfnamefont
  {S.}~\bibnamefont {De~Franceschi}},\ }\href {\doibase
  10.1103/PhysRevB.95.180502} {\bibfield  {journal} {\bibinfo  {journal} {Phys.
  Rev. B}\ }\textbf {\bibinfo {volume} {95}},\ \bibinfo {pages} {180502}
  (\bibinfo {year} {2017})}\BibitemShut {NoStop}%
\bibitem [{\citenamefont {Su}\ \emph {et~al.}(2017)\citenamefont {Su},
  \citenamefont {Tacla}, \citenamefont {Hocevar}, \citenamefont {Car},
  \citenamefont {Plissard}, \citenamefont {Bakkers}, \citenamefont {Daley},
  \citenamefont {Pekker},\ and\ \citenamefont {Frolov}}]{su_andreevmolecule}%
  \BibitemOpen
  \bibfield  {author} {\bibinfo {author} {\bibfnamefont {Z.}~\bibnamefont
  {Su}}, \bibinfo {author} {\bibfnamefont {A.~B.}\ \bibnamefont {Tacla}},
  \bibinfo {author} {\bibfnamefont {M.}~\bibnamefont {Hocevar}}, \bibinfo
  {author} {\bibfnamefont {D.}~\bibnamefont {Car}}, \bibinfo {author}
  {\bibfnamefont {S.~R.}\ \bibnamefont {Plissard}}, \bibinfo {author}
  {\bibfnamefont {E.~P. A.~M.}\ \bibnamefont {Bakkers}}, \bibinfo {author}
  {\bibfnamefont {A.~J.}\ \bibnamefont {Daley}}, \bibinfo {author}
  {\bibfnamefont {D.}~\bibnamefont {Pekker}}, \ and\ \bibinfo {author}
  {\bibfnamefont {S.~M.}\ \bibnamefont {Frolov}},\ }\href@noop {} {\bibfield
  {journal} {\bibinfo  {journal} {Nature Communications}\ }\textbf {\bibinfo
  {volume} {8}},\ \bibinfo {pages} {585} (\bibinfo {year} {2017})}\BibitemShut
  {NoStop}%
\bibitem [{\citenamefont {Su}\ \emph {et~al.}(2018)\citenamefont {Su},
  \citenamefont {Zarassi}, \citenamefont {Hsu}, \citenamefont {San-Jose},
  \citenamefont {Prada}, \citenamefont {Aguado}, \citenamefont {Lee},
  \citenamefont {Gazibegovic}, \citenamefont {Op~het Veld}, \citenamefont
  {Car}, \citenamefont {Plissard}, \citenamefont {Hocevar}, \citenamefont
  {Pendharkar}, \citenamefont {Lee}, \citenamefont {Logan}, \citenamefont
  {Palmstr\o{}m}, \citenamefont {Bakkers},\ and\ \citenamefont
  {Frolov}}]{SuPRL18}%
  \BibitemOpen
  \bibfield  {author} {\bibinfo {author} {\bibfnamefont {Z.}~\bibnamefont
  {Su}}, \bibinfo {author} {\bibfnamefont {A.}~\bibnamefont {Zarassi}},
  \bibinfo {author} {\bibfnamefont {J.-F.}\ \bibnamefont {Hsu}}, \bibinfo
  {author} {\bibfnamefont {P.}~\bibnamefont {San-Jose}}, \bibinfo {author}
  {\bibfnamefont {E.}~\bibnamefont {Prada}}, \bibinfo {author} {\bibfnamefont
  {R.}~\bibnamefont {Aguado}}, \bibinfo {author} {\bibfnamefont {E.~J.~H.}\
  \bibnamefont {Lee}}, \bibinfo {author} {\bibfnamefont {S.}~\bibnamefont
  {Gazibegovic}}, \bibinfo {author} {\bibfnamefont {R.~L.~M.}\ \bibnamefont
  {Op~het Veld}}, \bibinfo {author} {\bibfnamefont {D.}~\bibnamefont {Car}},
  \bibinfo {author} {\bibfnamefont {S.~R.}\ \bibnamefont {Plissard}}, \bibinfo
  {author} {\bibfnamefont {M.}~\bibnamefont {Hocevar}}, \bibinfo {author}
  {\bibfnamefont {M.}~\bibnamefont {Pendharkar}}, \bibinfo {author}
  {\bibfnamefont {J.~S.}\ \bibnamefont {Lee}}, \bibinfo {author} {\bibfnamefont
  {J.~A.}\ \bibnamefont {Logan}}, \bibinfo {author} {\bibfnamefont {C.~J.}\
  \bibnamefont {Palmstr\o{}m}}, \bibinfo {author} {\bibfnamefont {E.~P. A.~M.}\
  \bibnamefont {Bakkers}}, \ and\ \bibinfo {author} {\bibfnamefont {S.~M.}\
  \bibnamefont {Frolov}},\ }\href {\doibase 10.1103/PhysRevLett.121.127705}
  {\bibfield  {journal} {\bibinfo  {journal} {Phys. Rev. Lett.}\ }\textbf
  {\bibinfo {volume} {121}},\ \bibinfo {pages} {127705} (\bibinfo {year}
  {2018})}\BibitemShut {NoStop}%
\bibitem [{\citenamefont {Steele}\ \emph {et~al.}()\citenamefont {Steele},
  \citenamefont {G\"otz},\ and\ \citenamefont
  {Kouwenhoven}}]{steele_unpublished}%
  \BibitemOpen
  \bibfield  {author} {\bibinfo {author} {\bibfnamefont {G.~A.}\ \bibnamefont
  {Steele}}, \bibinfo {author} {\bibfnamefont {G.}~\bibnamefont {G\"otz}}, \
  and\ \bibinfo {author} {\bibfnamefont {L.~P.}\ \bibnamefont {Kouwenhoven}},\
  }\href@noop {} {\bibinfo  {journal} {unpublished}\ }\BibitemShut {NoStop}%
\bibitem [{\citenamefont {Mart{\'\i}n-Rodero}\ and\ \citenamefont
  {Levy~Yeyati}(2011)}]{martin2011josephson}%
  \BibitemOpen
\bibfield  {journal} {  }\bibfield  {author} {\bibinfo {author} {\bibfnamefont
  {A.}~\bibnamefont {Mart{\'\i}n-Rodero}}\ and\ \bibinfo {author}
  {\bibfnamefont {A.}~\bibnamefont {Levy~Yeyati}},\ }\href@noop {} {\bibfield
  {journal} {\bibinfo  {journal} {Advances in Physics}\ }\textbf {\bibinfo
  {volume} {60}},\ \bibinfo {pages} {899} (\bibinfo {year} {2011})}\BibitemShut
  {NoStop}%
\bibitem [{\citenamefont {Luitz}\ \emph {et~al.}(2012)\citenamefont {Luitz},
  \citenamefont {Assaad}, \citenamefont {Novotn{\'{y}}}, \citenamefont
  {Karrasch},\ and\ \citenamefont {Meden}}]{Luitz:2012jz}%
  \BibitemOpen
  \bibfield  {author} {\bibinfo {author} {\bibfnamefont {D.~J.}\ \bibnamefont
  {Luitz}}, \bibinfo {author} {\bibfnamefont {F.~F.}\ \bibnamefont {Assaad}},
  \bibinfo {author} {\bibfnamefont {T.}~\bibnamefont {Novotn{\'{y}}}}, \bibinfo
  {author} {\bibfnamefont {C.}~\bibnamefont {Karrasch}}, \ and\ \bibinfo
  {author} {\bibfnamefont {V.}~\bibnamefont {Meden}},\ }\href@noop {}
  {\bibfield  {journal} {\bibinfo  {journal} {Physical Review Letters}\
  }\textbf {\bibinfo {volume} {108}},\ \bibinfo {pages} {227001} (\bibinfo
  {year} {2012})}\BibitemShut {NoStop}%
\bibitem [{\citenamefont {Meden}(2018)}]{meden2018}%
  \BibitemOpen
  \bibfield  {author} {\bibinfo {author} {\bibfnamefont {V.}~\bibnamefont
  {Meden}},\ }\href@noop {} {\enquote {\bibinfo {title} {The
  {Anderson}-{Josephson} quantum dot -- a theory perspective},}\ }\bibinfo
  {howpublished} {arxiv:1810.02181} (\bibinfo {year} {2018})\BibitemShut
  {NoStop}%
\bibitem [{\citenamefont {Wilson}(1975)}]{wilson1975}%
  \BibitemOpen
  \bibfield  {author} {\bibinfo {author} {\bibfnamefont {K.~G.}\ \bibnamefont
  {Wilson}},\ }\href@noop {} {\bibfield  {journal} {\bibinfo  {journal} {Rev.
  Mod. Phys.}\ }\textbf {\bibinfo {volume} {47}},\ \bibinfo {pages} {773}
  (\bibinfo {year} {1975})}\BibitemShut {NoStop}%
\bibitem [{\citenamefont {Bulla}\ \emph {et~al.}(2008)\citenamefont {Bulla},
  \citenamefont {Costi},\ and\ \citenamefont {Pruschke}}]{bulla2008}%
  \BibitemOpen
  \bibfield  {author} {\bibinfo {author} {\bibfnamefont {R.}~\bibnamefont
  {Bulla}}, \bibinfo {author} {\bibfnamefont {T.}~\bibnamefont {Costi}}, \ and\
  \bibinfo {author} {\bibfnamefont {T.}~\bibnamefont {Pruschke}},\ }\href@noop
  {} {\bibfield  {journal} {\bibinfo  {journal} {Rev. Mod. Phys.}\ }\textbf
  {\bibinfo {volume} {80}},\ \bibinfo {pages} {395} (\bibinfo {year}
  {2008})}\BibitemShut {NoStop}%
\bibitem [{\citenamefont {Pillet}\ \emph {et~al.}(2013)\citenamefont {Pillet},
  \citenamefont {Joyez}, \citenamefont {\ifmmode~\check{Z}\else
  \v{Z}\fi{}itko},\ and\ \citenamefont {Goffman}}]{pillet_prb2013}%
  \BibitemOpen
  \bibfield  {author} {\bibinfo {author} {\bibfnamefont {J.-D.}\ \bibnamefont
  {Pillet}}, \bibinfo {author} {\bibfnamefont {P.}~\bibnamefont {Joyez}},
  \bibinfo {author} {\bibfnamefont {R.}~\bibnamefont {\ifmmode~\check{Z}\else
  \v{Z}\fi{}itko}}, \ and\ \bibinfo {author} {\bibfnamefont {M.~F.}\
  \bibnamefont {Goffman}},\ }\href {\doibase 10.1103/PhysRevB.88.045101}
  {\bibfield  {journal} {\bibinfo  {journal} {Phys. Rev. B}\ }\textbf {\bibinfo
  {volume} {88}},\ \bibinfo {pages} {045101} (\bibinfo {year}
  {2013})}\BibitemShut {NoStop}%
\bibitem [{\citenamefont {\v{Z}itko}(2009)}]{odesolv}%
  \BibitemOpen
  \bibfield  {author} {\bibinfo {author} {\bibfnamefont {R.}~\bibnamefont
  {\v{Z}itko}},\ }\href@noop {} {\bibfield  {journal} {\bibinfo  {journal}
  {Comp. Phys. Comm.}\ }\textbf {\bibinfo {volume} {180}},\ \bibinfo {pages}
  {1271} (\bibinfo {year} {2009})}\BibitemShut {NoStop}%
\bibitem [{\citenamefont {Jin-Guo~Liu}(2016)}]{liu2016}%
  \BibitemOpen
  \bibfield  {author} {\bibinfo {author} {\bibfnamefont {Q.-H.~W.}\
  \bibnamefont {Jin-Guo~Liu}, \bibfnamefont {Da~Wang}},\ }\href@noop {}
  {\bibfield  {journal} {\bibinfo  {journal} {Phys. Rev. B}\ }\textbf {\bibinfo
  {volume} {93}},\ \bibinfo {pages} {035102} (\bibinfo {year}
  {2016})}\BibitemShut {NoStop}%
\bibitem [{\citenamefont {Lutchyn}\ \emph {et~al.}(2010)\citenamefont
  {Lutchyn}, \citenamefont {Sau},\ and\ \citenamefont
  {Das~Sarma}}]{LutchynPRL2010}%
  \BibitemOpen
  \bibfield  {author} {\bibinfo {author} {\bibfnamefont {R.~M.}\ \bibnamefont
  {Lutchyn}}, \bibinfo {author} {\bibfnamefont {J.~D.}\ \bibnamefont {Sau}}, \
  and\ \bibinfo {author} {\bibfnamefont {S.}~\bibnamefont {Das~Sarma}},\
  }\href@noop {} {\bibfield  {journal} {\bibinfo  {journal} {Physical review
  letters}\ }\textbf {\bibinfo {volume} {105}},\ \bibinfo {pages} {077001}
  (\bibinfo {year} {2010})}\BibitemShut {NoStop}%
\bibitem [{\citenamefont {Oreg}\ \emph {et~al.}(2010)\citenamefont {Oreg},
  \citenamefont {Refael},\ and\ \citenamefont {von Oppen}}]{OregPRL2010}%
  \BibitemOpen
  \bibfield  {author} {\bibinfo {author} {\bibfnamefont {Y.}~\bibnamefont
  {Oreg}}, \bibinfo {author} {\bibfnamefont {G.}~\bibnamefont {Refael}}, \ and\
  \bibinfo {author} {\bibfnamefont {F.}~\bibnamefont {von Oppen}},\ }\href@noop
  {} {\bibfield  {journal} {\bibinfo  {journal} {Physical review letters}\
  }\textbf {\bibinfo {volume} {105}},\ \bibinfo {pages} {177002} (\bibinfo
  {year} {2010})}\BibitemShut {NoStop}%
\end{thebibliography}%

\section{SUPPLEMENTAL MATERIALS}
\setcounter{figure}{0}
\renewcommand{\thefigure}{S\arabic{figure}}

\begin{figure*}[t!]
\centering
\includegraphics[width=0.8\textwidth]{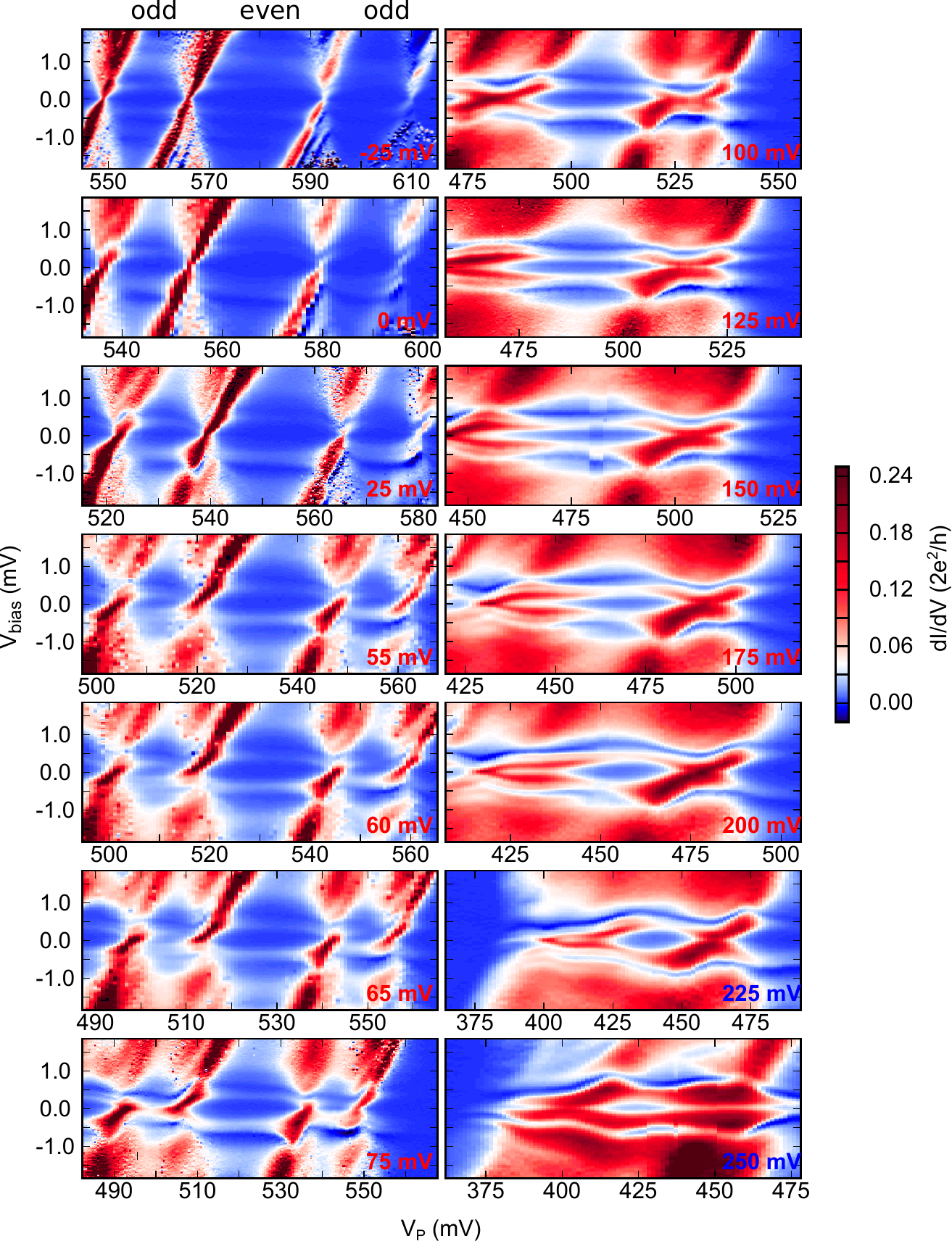}
\caption{\textbf{Full gate tunability from weak to strong coupling between a quantum dot and a superconductor}. Andreev bound states with different superconducting coupling strengths in sample studied in the main text. The coupling is tuned by the gate voltage $V_S$ which is noted in each panel in the bottom right. The data start in the co-tunneling regime with well-resolved Coulomb diamonds in the upper-left panel. As $V_S$ is made more positive, the coupling to the superconductor gets stronger and the Andreev bound states within the odd-parity Coulomb diamonds transition from spin-doublet to spin-singlet ground state. As a result, loops of the odd parity Andreev bound states shrink and finally disappear. At the highest $V_S$=250 mV, the ground state is always spin-singlet (even parity). At the same time, above the gap, the features of sharp Coulomb diamonds are replaced by broad resonances with double the period in gate voltage.}
\end{figure*}

\newpage

\begin{figure*}[t!]
\centering
\includegraphics[width=\textwidth]{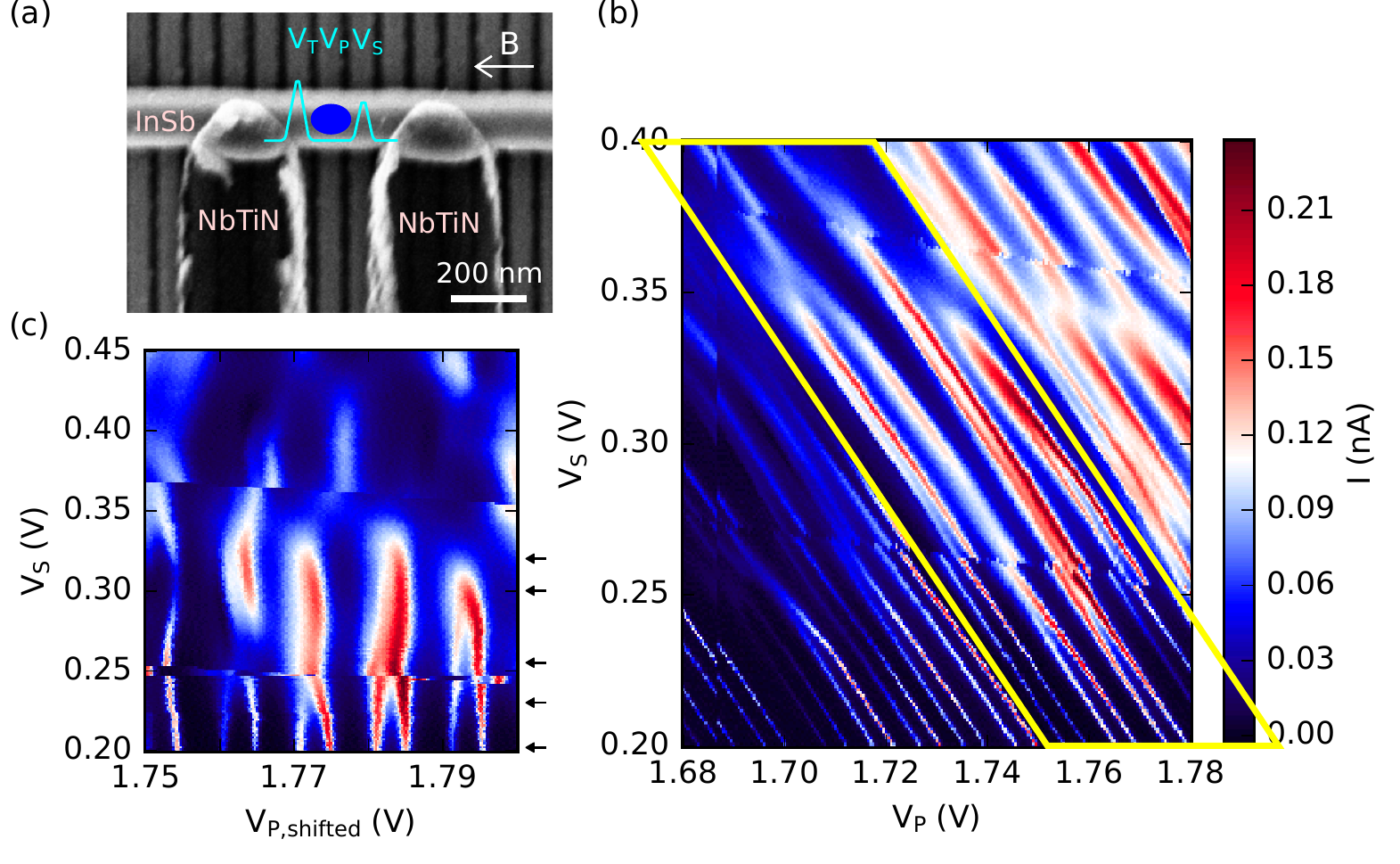}
\caption{\textbf{Measurements from another device.} (a) Scanning electron microscope picture of the second device (only presented in supplemental materials). $V_T$, $V_P$ and $V_S$ are gate voltages tuning the tunneling barrier, the chemical potential and the superconducting coupling of the dot. The InSb nanowire is grown by metalorganic vapour-phase epitaxy and transferred to the substrate by a micro-manipulator. NbTiN superconducting contacts are sputtered to induce superconductivity in the nanowire. We note that even though the tunnel probe is a superconductor, soft induced gap in the nanowire creates a finite density of states at the Fermi level, making it possible to perform low-bias (subgap) spectroscopy by directly reading the positions of energy levels from the source-drain bias voltage, without subtracting a gap. Measurements are performed in a dry dilution fridge at a base temperature of 40 mK. The direction of the magnetic field is parallel to the nanowire. (b) $V_S$ - $V_P$ diagram for the second device, at a source-drain bias of 30 $\mu$V. A series of Coulomb peaks are seen at $V_S$ = 0.2 V, which merge pairwise at higher $V_S$. The discontinuances around $V_S$ = 0.27 V and 0.37 V is an artificial effect due to charge jumps in a spurious quantum dot capacitively coupled to the quantum dot under investigation. (c) In order to compensate the shift of features in panel (b) due to direct coupling of $V_S$ to the quantum dot states, we perform "shifted-scans" in the area enclosed by the yellow parallelogram of panel (b). The horizontal axis in panel (c) thus corresponds to a range of $V_P$ shifted for each $V_S$. Panel (c) shows more clearly the pairwise merger of Coulomb peaks at more positive $V_S$. The merged peaks vanish at high $V_S$ due to low applied source-drain bias. Eventually the ABS resonance moves to higher bias in the strongly coupled regime. The magnetic field evolution at different $V_S$ marked by the arrows is show in Fig. \ref{B_vs_Vp}.}
\label{S2}
\end{figure*}

\newpage

\begin{figure*}
\centering
\includegraphics[width=\textwidth]{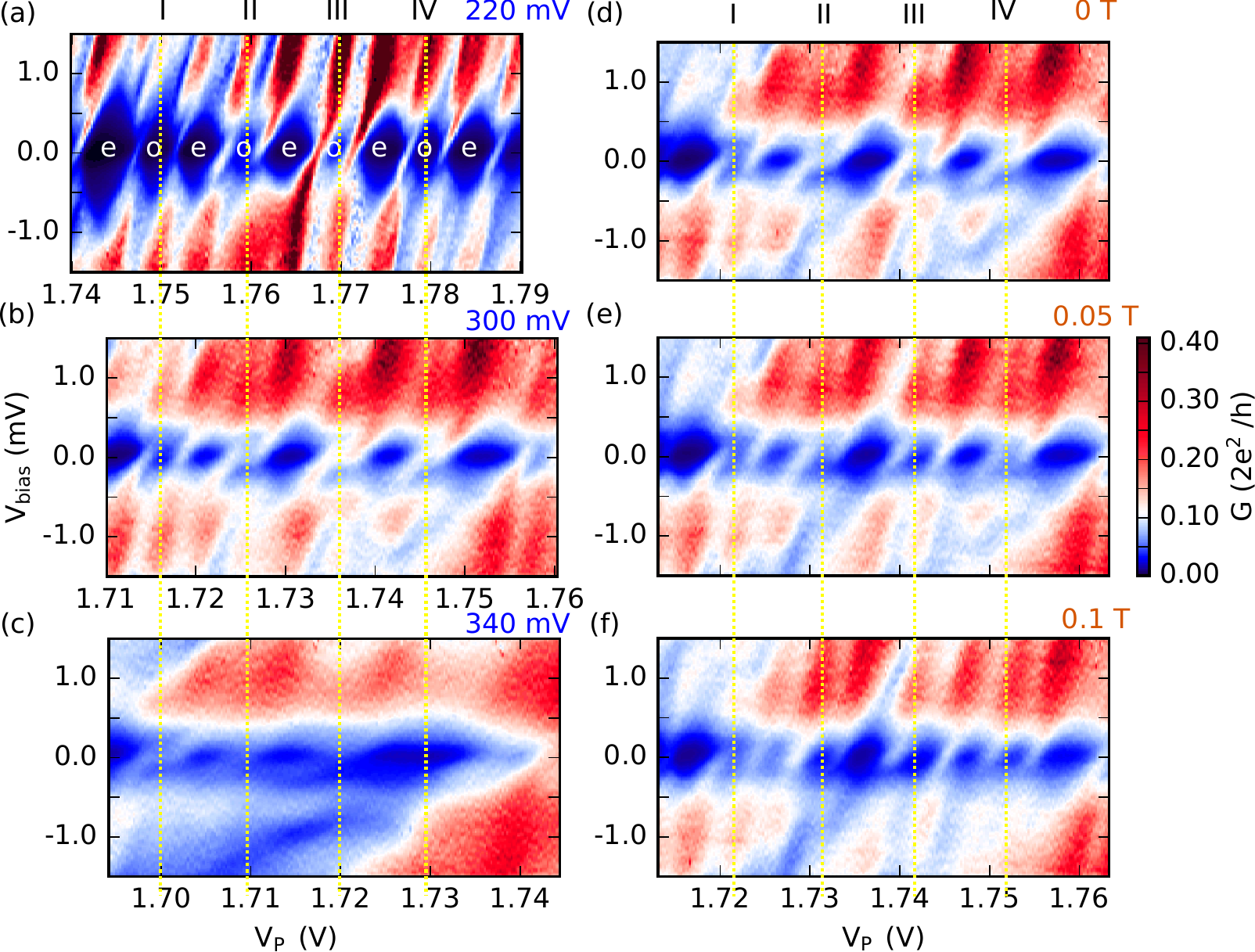}
\caption{(a)-(c) V$_{bias}$ vs $V_P$ at different superconducting couplings in the second device. $V_S$ is noted in each panel in the upper-right corner. (a) Coulomb diamonds with even-odd periodicity can be observed when $V_S$ = 220 mV, in the weak coupling regime. (b) When $V_S$ = 300 mV, diamonds are no longer clearly resolved, instead conductance increases above $V_{bias}=0.5$meV, the induced gap. Faint in-gap states can be identified as loop-like ABS resonances. (c) When $V_S$ is 340 mV, the strongest coupling to the superconductor, the high-bias conductance periodicity is larger (double) than in panel (a). At low bias, faint anticrossing-like resonances can be observed, within the regime labeled II, III. (d)-(f) Zeeman splitting of Andreev bound states at $V_S$ = 300 mV which corresponds to panel (b). }
\end{figure*}

\newpage

\begin{figure*}[t!]
\centering
\includegraphics[width=\textwidth]{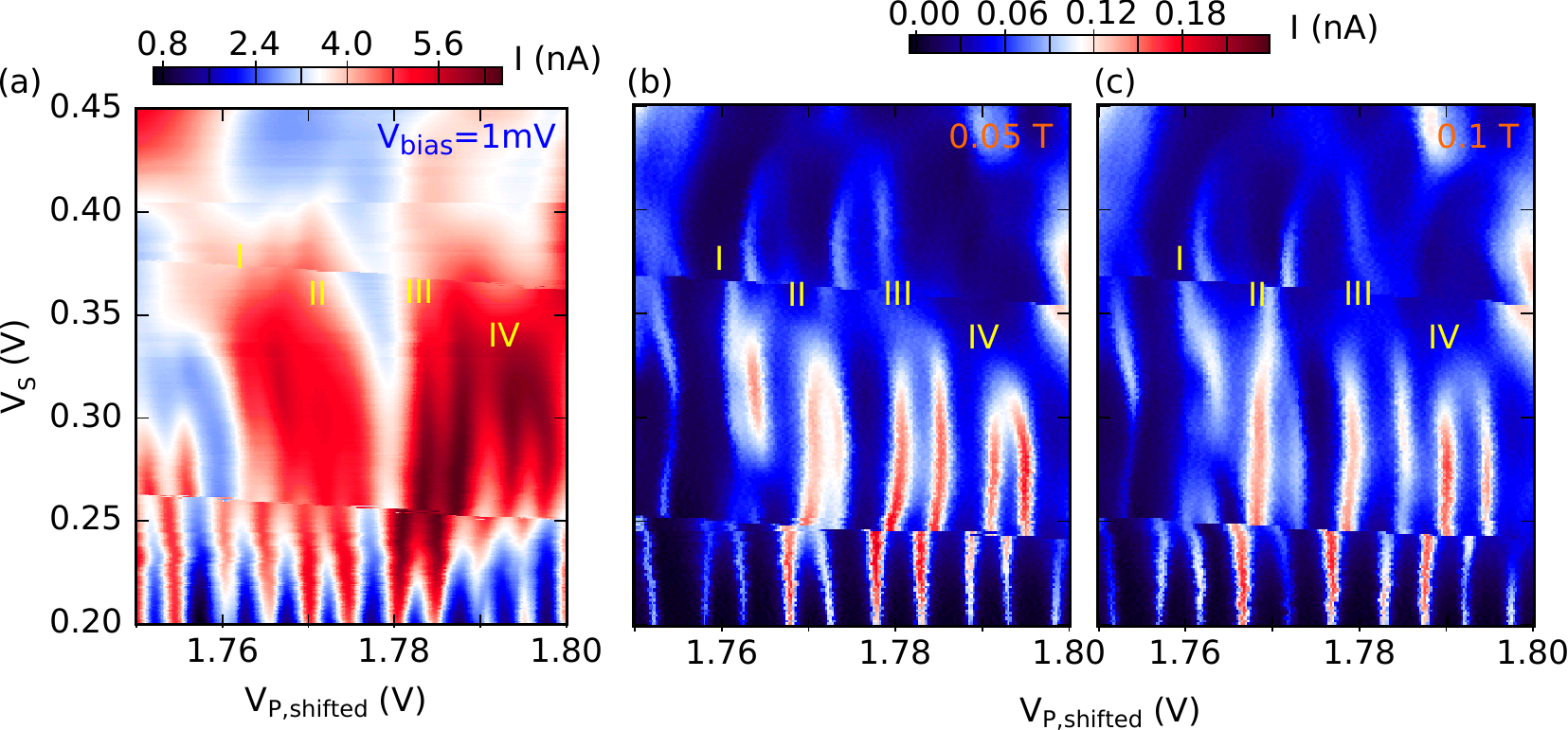}
\caption{Conductance maps at (a) above-gap source-drain bias (zero magnetic field) and (b) (c) different magnetic fields (30 $\mu$V bias voltage) taken from the same regime as Fig. \ref{S2}(c), in the second device. The bias voltage or magnetic field are noted in each panel.}
\end{figure*}

\newpage

\begin{figure*}[t!]
\centering
\includegraphics[width=0.5\textwidth]{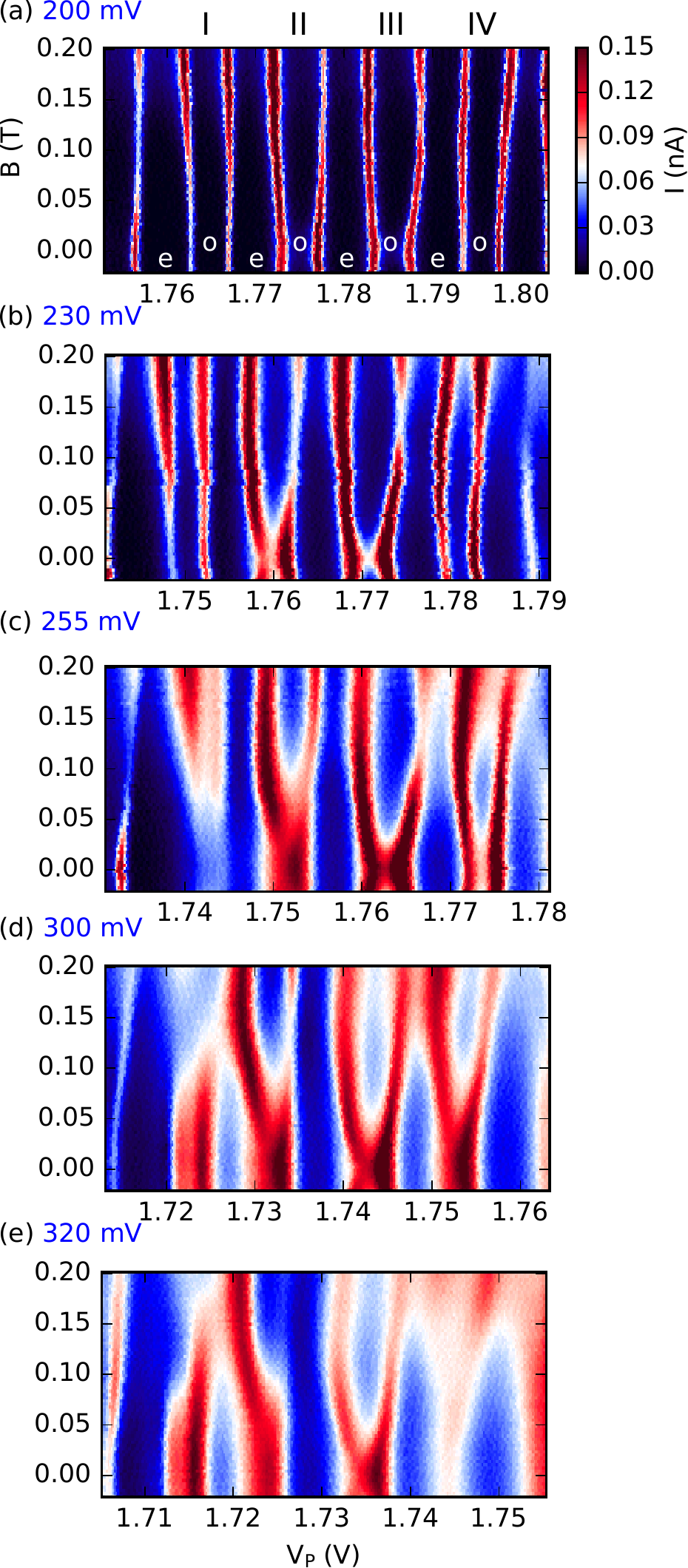}  
\caption{Magnetic field evolution of plunger gate at different $V_S$ marked by arrows in Fig. \ref{S2}(c)}
\label{B_vs_Vp}
\end{figure*}

\clearpage

\begin{figure*}
\centering
\includegraphics[width=\textwidth]{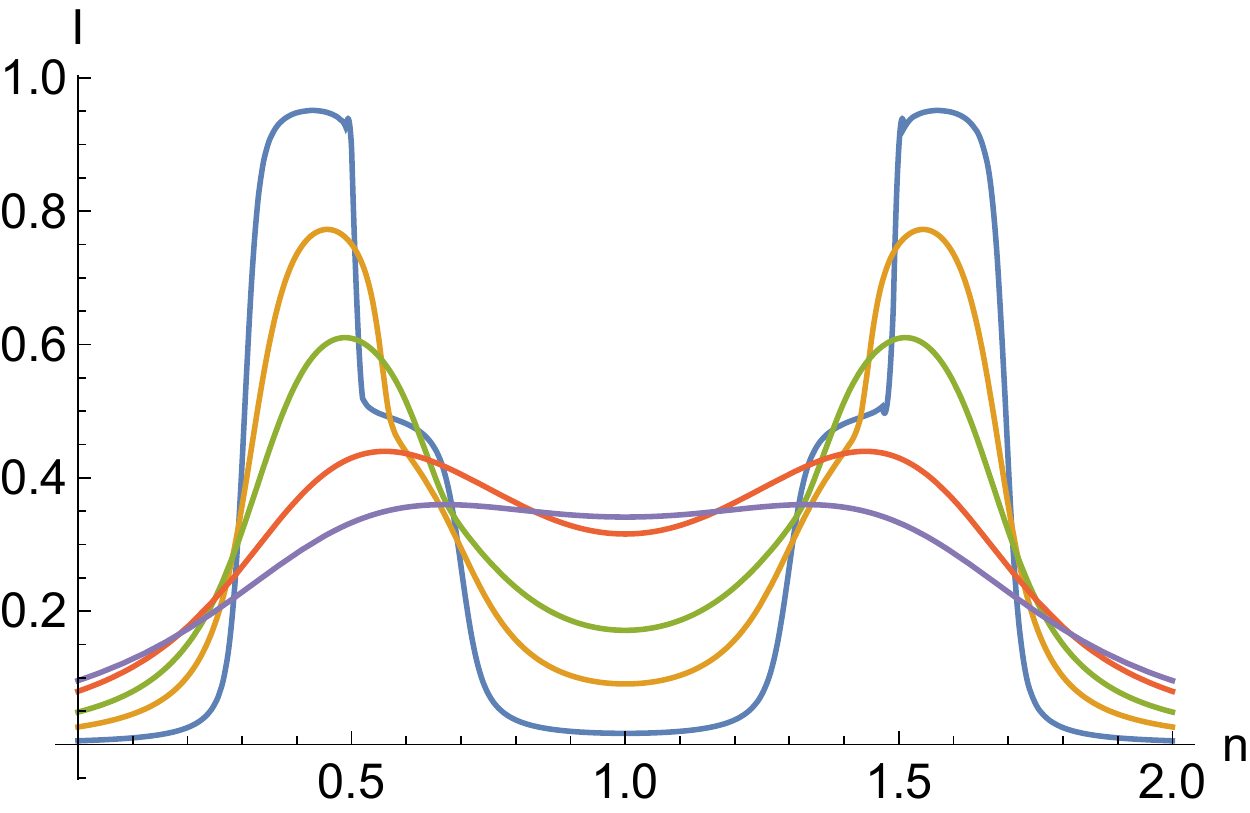}
\caption{Simulated finite-bias current vs. gate voltage (in units of charge) for a range of hybridization strengths $\Gamma$: 0.01, 0.05, 0.1, 0.2 and 0.3. These are horizontal cuts through the density plot in Fig.~4(a) on the main text.}
\label{rz1}
\end{figure*}


\begin{figure*}
\centering
\includegraphics[width=0.8\textwidth]{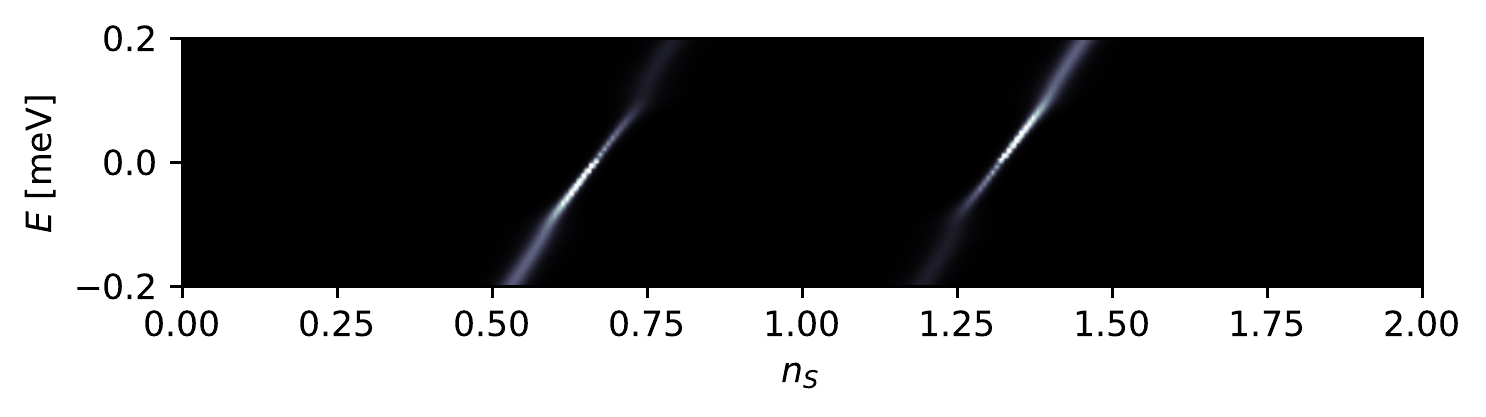}\\
\includegraphics[width=0.8\textwidth]{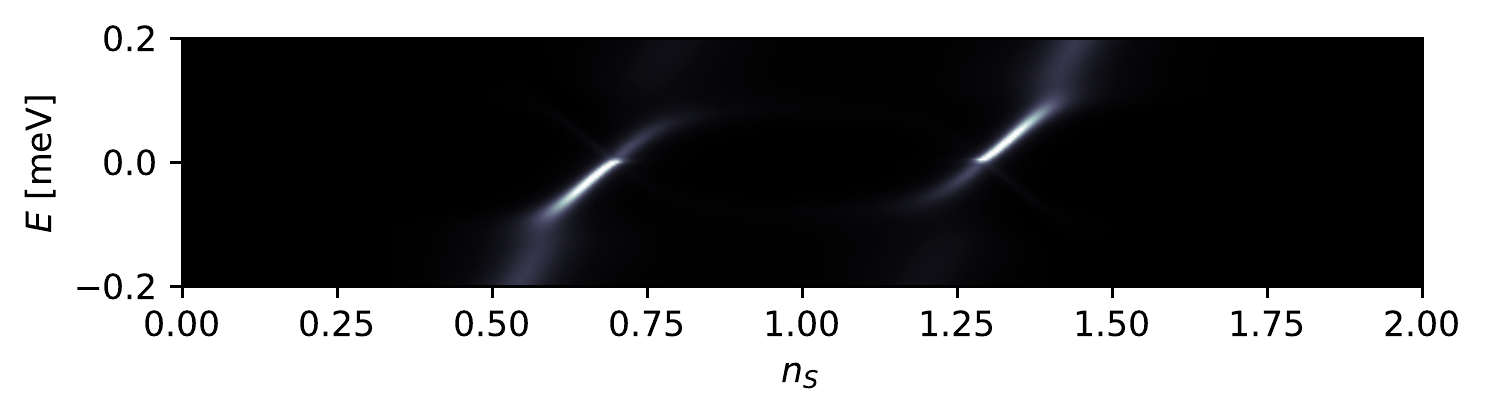}\\
\includegraphics[width=0.8\textwidth]{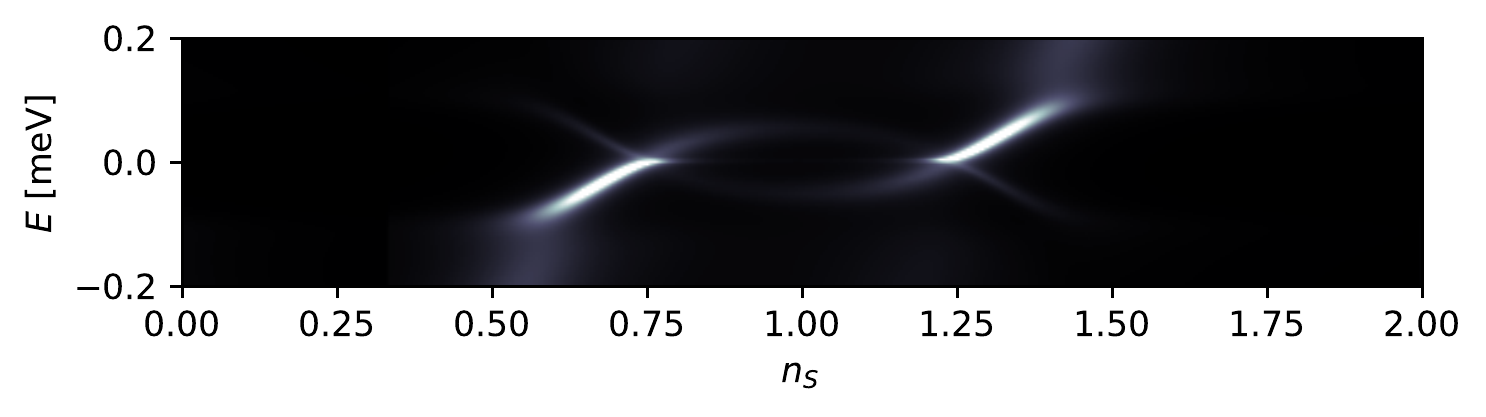}\\
\includegraphics[width=0.8\textwidth]{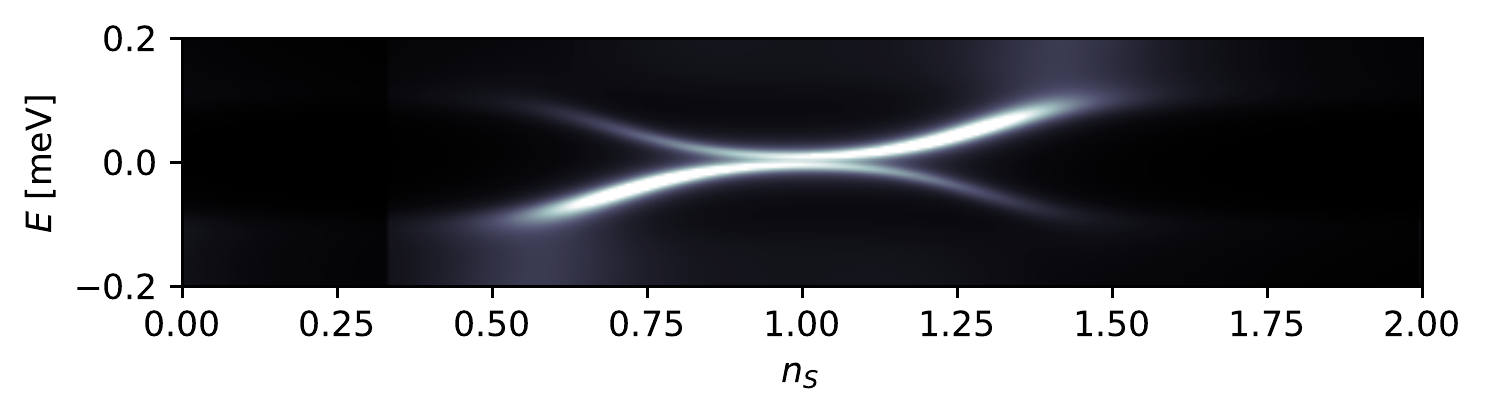}\\
\includegraphics[width=0.8\textwidth]{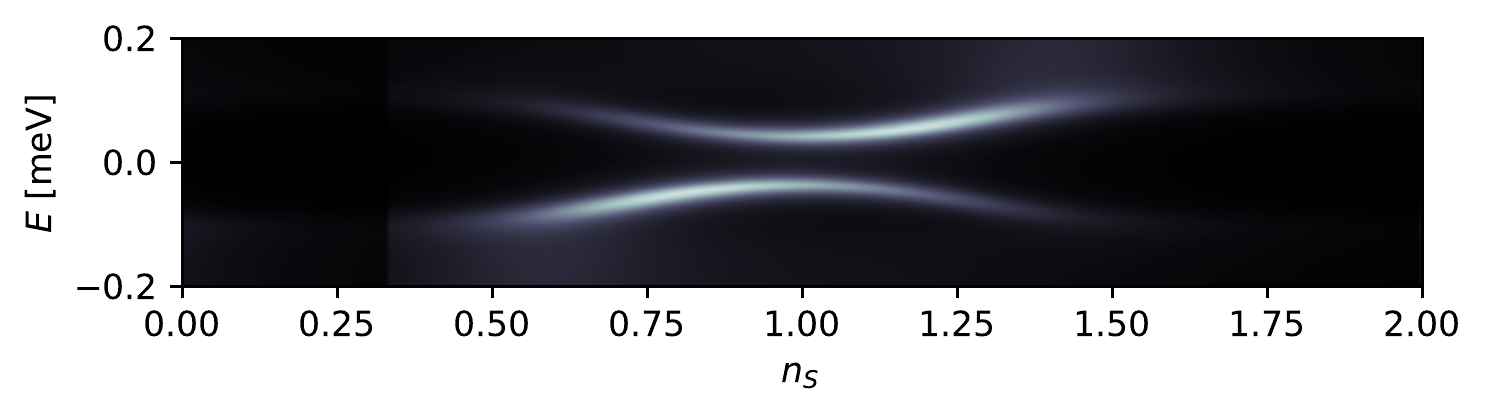}
\caption{Spectral functions that correspond to the cuts shown in preceding figure, Fig.~\ref{rz1}.}
\end{figure*}

\begin{figure*}
\centering
\includegraphics[width=0.4\textwidth]{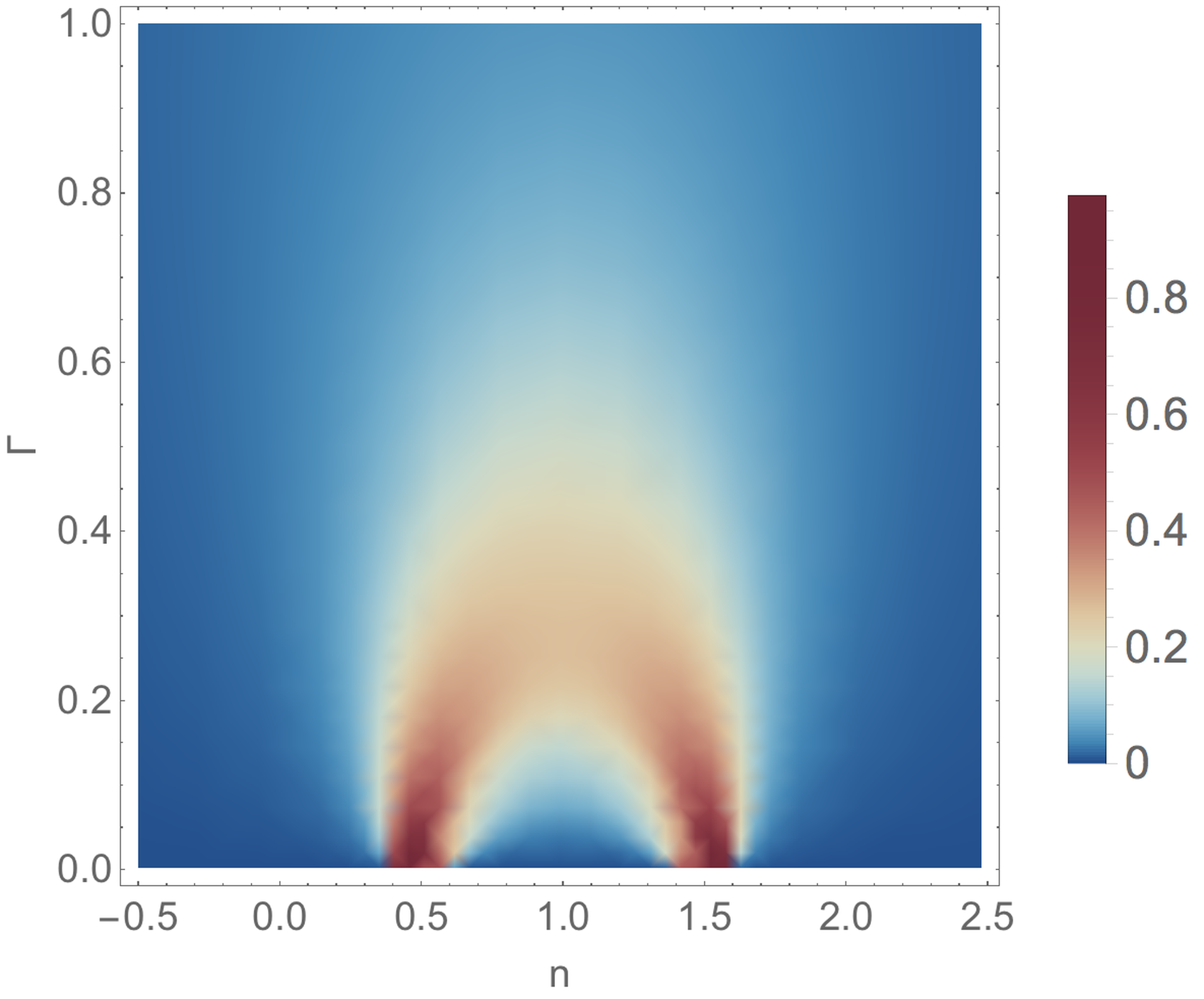}
\includegraphics[width=0.4\textwidth]{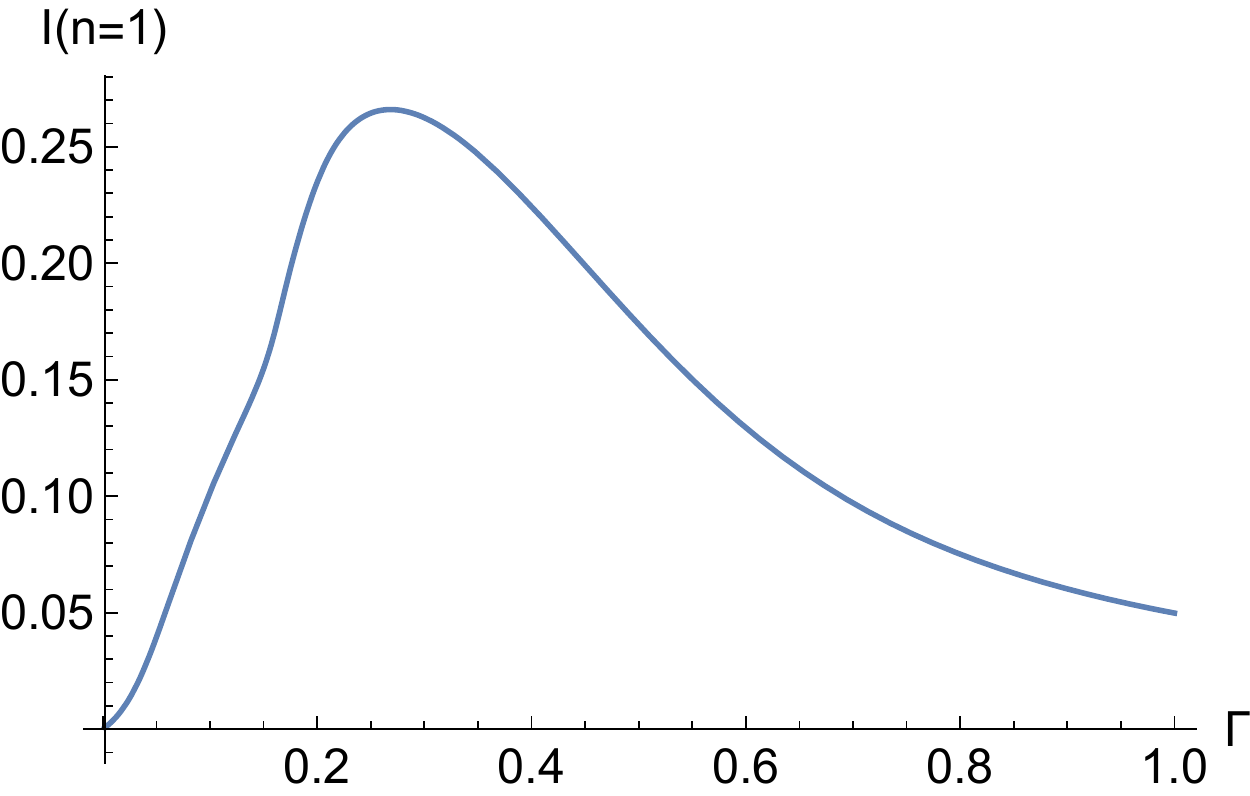}\\
\includegraphics[width=0.4\textwidth]{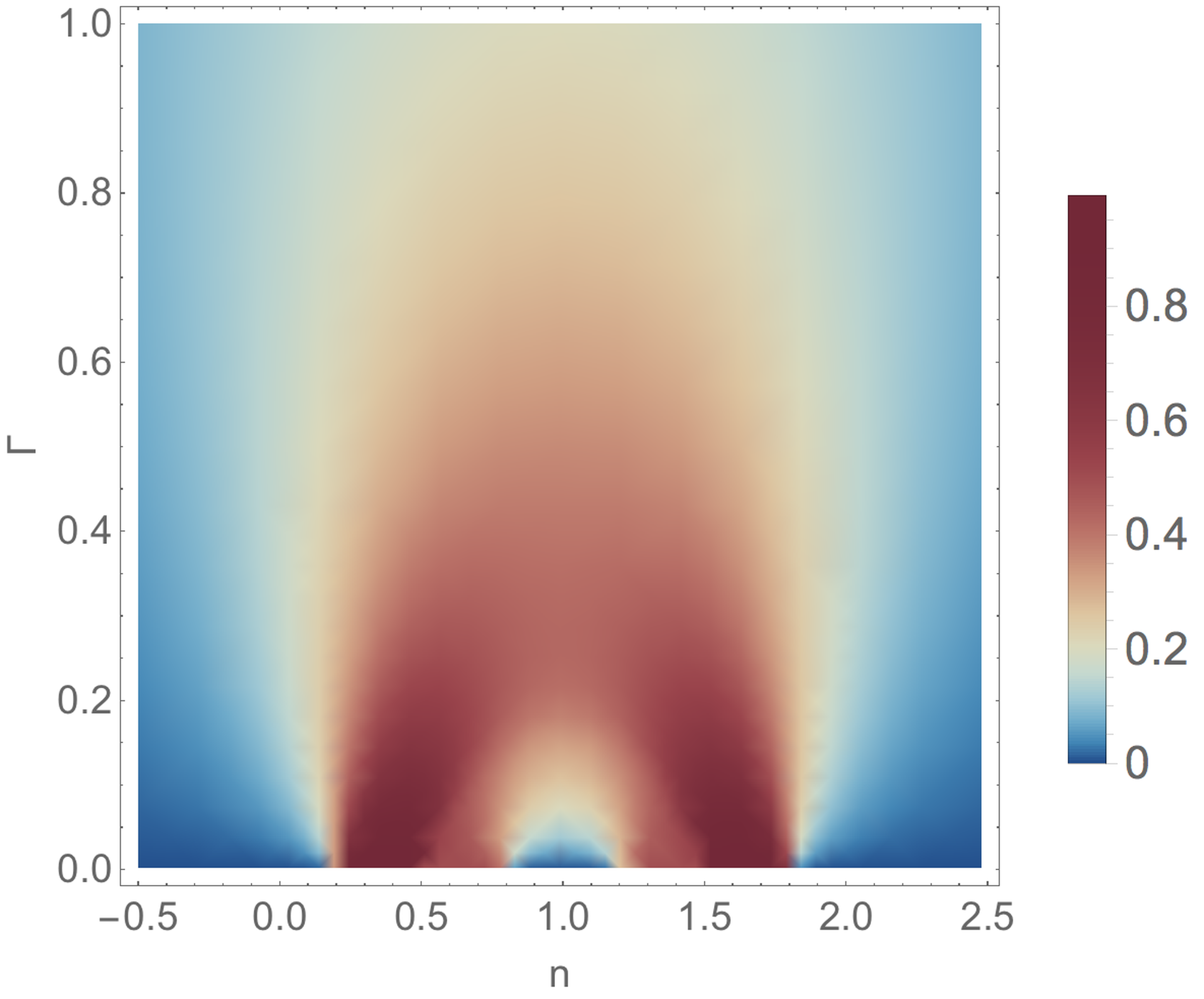}
\includegraphics[width=0.4\textwidth]{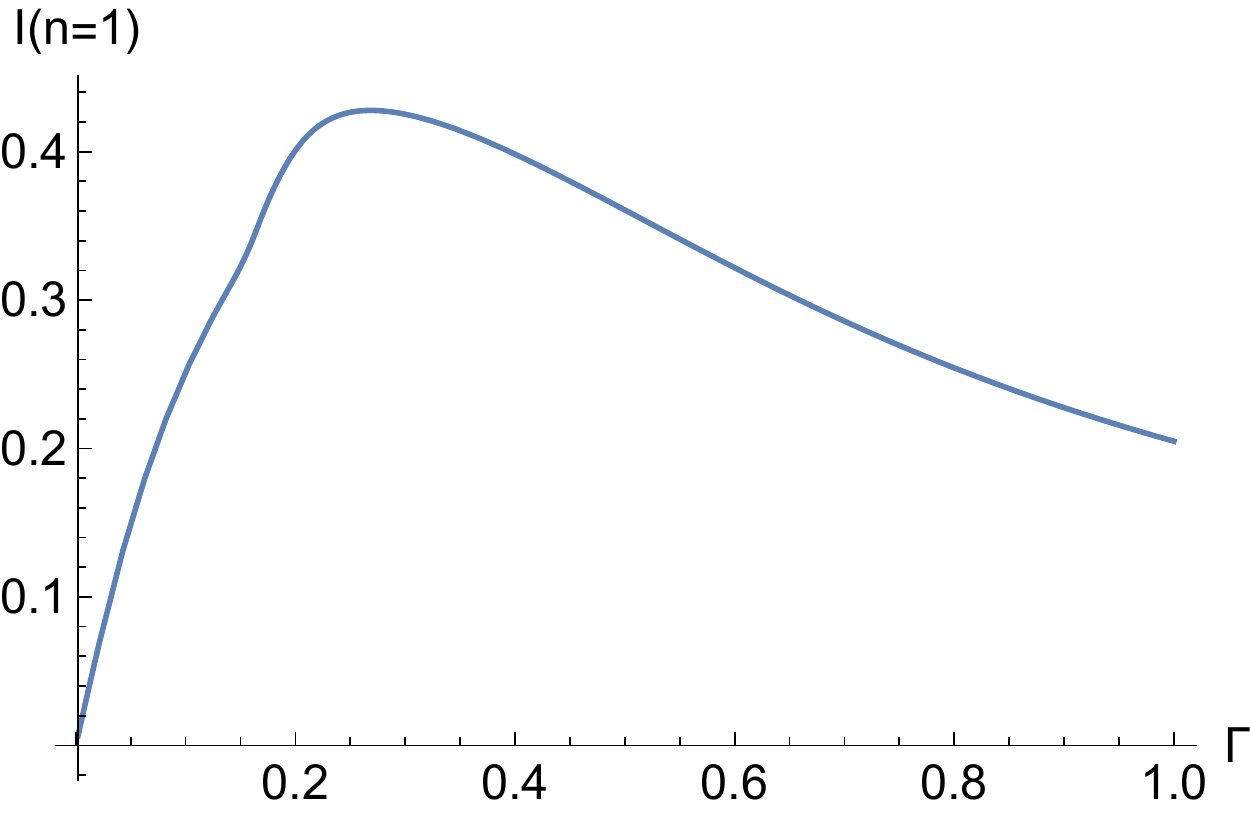}
\caption{Simulated finite-bias current as a function of gate voltage and hybridization strength (left panels) and a vertical cut for $n=1$ (right panels), for different integration ranges, from $-\Delta$ to $\Delta$ (top panels), and from $-3\Delta$
to $3\Delta$ (bottom panels). This is to be compared with Fig.~4 in the main text, where the integration is performed from $-2\Delta$ to $2\Delta$. Note that the case of top panels represents the contribution from the sub-gap states only.}
\end{figure*}

\begin{figure*}
\centering
\includegraphics[width=0.98\textwidth]{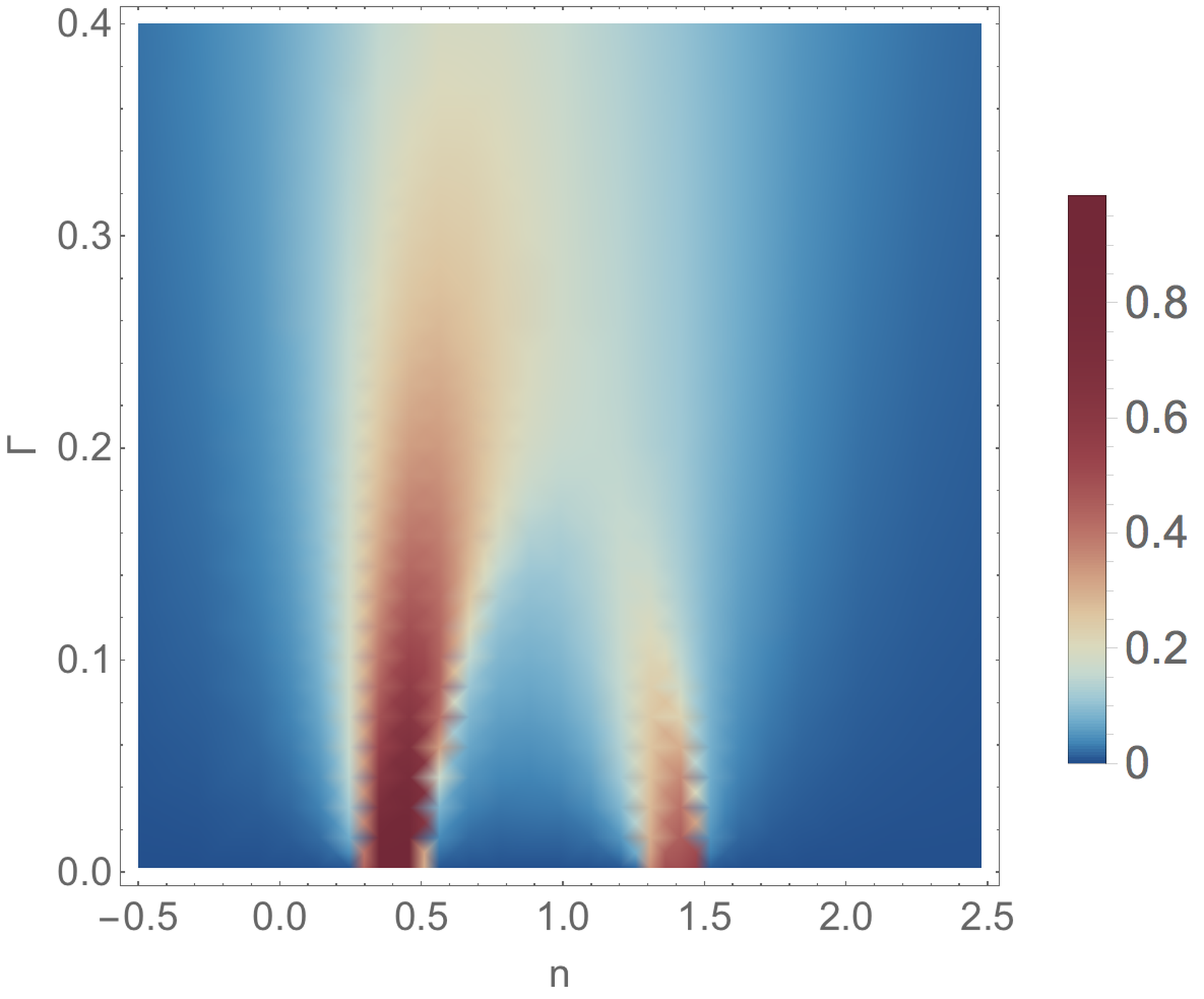}
\caption{Simulated finite-bias current as a function of gate voltage and hybridization strength for an asymmetric integration range, from 0 to $2\Delta$.}
\end{figure*}

\end{document}